\documentclass[preprint,showpacs,preprintnumbers,amsmath,amssymb, aps, pra, nofootinbib]{revtex4-2}  
\eprint
\usepackage{lipsum}
\usepackage[symbol]{footmisc}
\usepackage{graphicx}
\usepackage{amsmath}
\usepackage{tabularx}
\usepackage{mathtools}

\DeclarePairedDelimiterX\braket[2]{\langle}{\rangle}{#1 \delimsize\vert #2}
\usepackage{bm}
\usepackage{color}
\usepackage{esvect}
\usepackage{gensymb}
\usepackage{parskip}

\usepackage{ulem}

\usepackage{comment}
\usepackage[colorlinks=true, citecolor=blue, linkcolor=blue, urlcolor=blue]{hyperref}
\usepackage{tikz,lipsum,lmodern}
\usepackage[most]{tcolorbox}
\usepackage{textcase} 
\usepackage{hyperref}
\usepackage{setspace}
\AtBeginDocument{\singlespacing} 
\usepackage{xcolor}
\hypersetup{
  colorlinks   = true, 
  urlcolor     = blue, 
  linkcolor    = blue, 
  citecolor   = blue 
}
\usepackage[all]{hypcap} 

\begin{document}

\title{Terahertz field-induced giant symmetry modulations in a van der Waals antiferromagnet}

\author{Sheikh Rubaiat Ul Haque$^{1,2,3,10}$}  \email{rubaiath@stanford.edu}
\author{Martin J. Cross$^{4,10}$}
\author{Sangeeta Rajpurohit$^{5,6,10}$}
\author{Jonah B. Haber$^{2,3,10}$}
\author{Christopher J. Ciccarino$^{2,3}$}
\author{Alexandra C. Zimmerman$^{2}$}
\author{Isabelle J. Sealey$^{7,8}$} 
\author{Vadym Kulichenko$^{7}$}
\author{Monique Tie$^{1,3}$}
\author{Huaiyu Wang$^{3}$}
\author{Sharon S. Philip$^{3}$}
\author{Choongwon Seo$^{4}$}
\author{Jake D. Koralek$^{4}$}
\author{Luis Balicas$^{7,8}$}
\author{Mykhaylo Ozerov$^{7}$}
\author{Dmitry Smirnov$^{7,8}$}
\author{Liang Z. Tan$^{6}$}
\author{Felipe H. da Jornada$^{2,3}$}
\author{Tadashi Ogitsu$^{5}$}
\author{Matthias C. Hoffmann$^{4}$} 
\author{Tony F. Heinz$^{1,3,9}$} \email{tony.heinz@stanford.edu}
\author{Aaron M. Lindenberg$^{2,3,9}$} \email{aaronl@stanford.edu}

\affiliation{
$^1$Department of Applied Physics, Stanford University, Stanford, CA 94305, USA\\
$^2$Department of Materials Science and Engineering, Stanford University, Stanford, CA 94305, USA\\
$^3$Stanford Institute for Materials and Energy Sciences, SLAC National Accelerator Laboratory, Menlo Park, CA 94025, USA\\
$^4$Linac Coherent Light Source, SLAC National Accelerator Laboratory, Menlo Park, California 94025, USA\\
$^5$Lawrence Livermore National Laboratory, Livermore, CA 94550, USA\\
$^6$Molecular Foundry, Lawrence Berkeley National Laboratory, Berkeley, CA 94720, USA\\
$^7$National High Magnetic Field Laboratory, Tallahassee, FL 32310, USA\\
$^8$Department of Physics, Florida State University, Tallahassee, FL 32306, USA\\
$^9$Stanford PULSE Institute, SLAC National Accelerator Laboratory, Menlo Park, CA 94025, USA\\
$^{10}$These authors contributed equally to the work}
 
\maketitle 

\noindent \textbf{Strong-field terahertz (THz) excitations enable dynamic control over electronic, lattice and symmetry degrees of freedom in quantum materials. Here, we uncover pronounced terahertz-induced symmetry modulations and coherent phonon dynamics in the van der Waals antiferromagnet MnPS$_3$, in which inversion symmetry is broken by its antiferromagnetic spin configuration. Time-resolved second harmonic generation measurements reveal long-lived giant oscillations in the antiferromagnetic phase, with amplitudes comparable to the equilibrium signal, driven by phonons involving percent-level atomic displacements relative to the equilibrium bond lengths. The temporal evolution of the rotational anisotropy  patterns indicate a dynamic breaking of mirror symmetry, modulated by two vibrational modes at 1.7 THz and 4.5 THz, with the former corresponding to a hidden mode not observed in equilibrium spectroscopy. We show that these effects arise in part from a field-induced charge rearrangement mechanism that lowers the local crystal symmetry, and couples to the phonon modes. A long-lived field-driven response was uncovered with a complex THz polarization dependence which, in comparison to theory, indicates evidence for an antiferromagnetic-to-ferrimagnetic transition. Our results establish an effective field-tunable pathway for driving excitations otherwise weak in equilibrium, and for manipulating magnetism in low-dimensional materials via dynamical modulation of symmetry.
}

Tailored intense laser pulses have emerged as a versatile tool to dynamically study quantum phenomena \cite{stojchevska2014, zhang2016, mitrano2016, Haque2024, kogar2020, wang2013} that are otherwise inaccessible by static tuning parameters such as pressure, voltage, strain, and temperature. Particularly, strong-field THz excitation of materials has attracted considerable interest for its unique ability to selectively drive low-energy collective modes. This includes THz-induced superconductivity \cite{rowe2023, budden2021}, ferroelectricity \cite{li2019, nova2019}, metal-insulator transition \cite{liu2012}, topological phase transition \cite{sie2019}, chirality \cite{zeng2025}, and ferromagnetism \cite{disa2020, disa2023}. Thus, harnessing coherent THz control could shed light on open questions in condensed matter and unlock a wide array of opportunities for next-generation photonics \cite{averitt2011, basov2017}.
 
\setcounter{figure}{0}
\renewcommand{\figurename}{\textbf{ Fig.}}
\renewcommand{\thefigure}{\arabic{figure}}
\begin{figure}
    \centering
    \includegraphics[width=\linewidth]{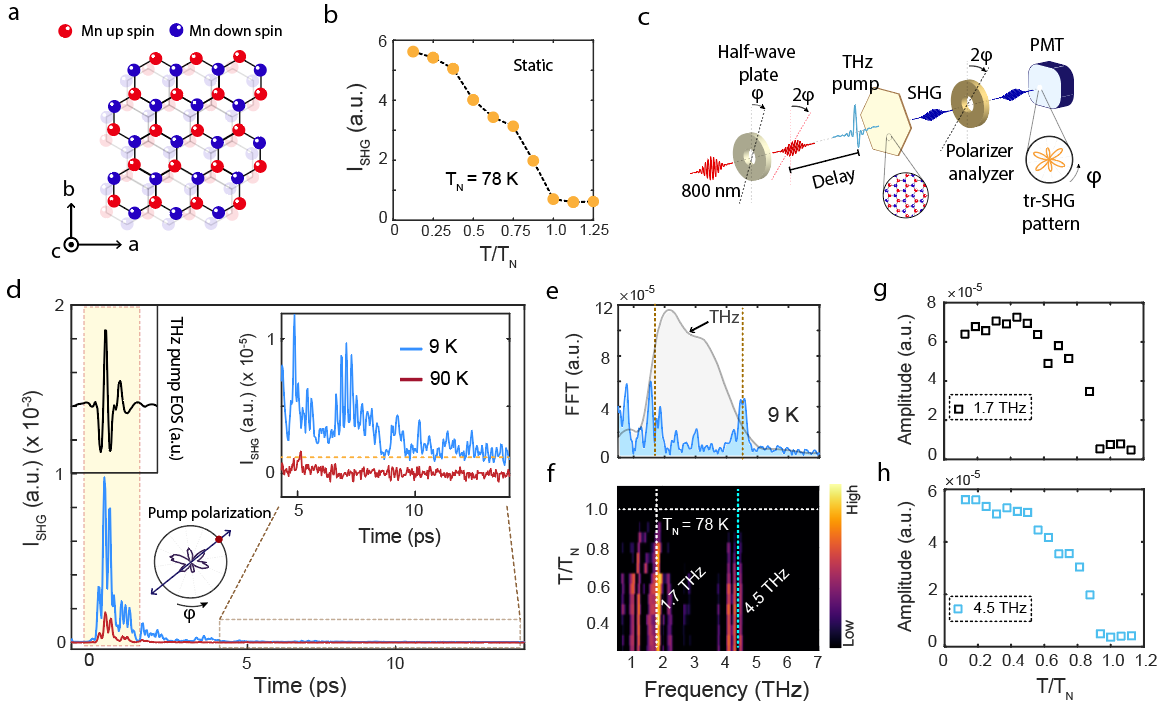}
    \caption{\textbf{MnPS$_3$ properies and THz pump-induced giant oscillations. a}, N\'eel-type layered AFM configuration in MnPS$_3$, blue and red spheres denote Mn atoms with opposite spins. \textbf{b}, Temperature-dependent SHG intensity measured along a lobe in the static SHG-RA pattern, exhibiting an onset below $T_N=78$ K. \textbf{c}, An intense terahertz pump pulse excites the sample and the induced changes are encoded in time-resolved (tr-) SHG-RA patterns produced by a delayed 800 nm probe pulse. \textbf{d}, THz-induced tr-SHG signals for temperatures below (9 K) and above (90 K) $T_\mathrm{N}$. The dark blue arrow indicates the pump polarization direction relative to the static SHG-RA pattern, and the red circle marks the angular position of the monitored signal. Yellow shaded region represents TFISH regime, with the THz pump profile temporally overlaid (black curve). The inset highlights the later-time dynamics, revealing giant oscillations at low temperatures with amplitude comparable to the static SHG signal. Horizontal orange dashed line denotes the equilibrium value. \textbf{e}, Amplitude spectrum of the oscillations at $T=9$ K, displaying prominent phonon peaks at 1.7 THz and 4.5 THz. The THz pump spectrum is shown in gray. \textbf{f}, Temperature-dependence of the spectrum. \textbf{g--h}, Both 1.7 THz (\textbf{g}) and 4.5 THz mode (\textbf{h}) amplitudes decrease with increasing temperature in a order parameter-like fashion. }
    \label{fig:1}
\end{figure}

A major theme in the study of correlated systems is magnetism, and advances in ultrafast laser sources facilitate optical manipulation of exotic magnetic phases on pico- or femtosecond timescales \cite{disa2020, disa2023, afanasiev2021, zong2023}. Antiferromagnets (AFM) are promising candidates for ultrafast spintronic applications since their zero net magnetization, originating from antiparallel spin configuration, enables much faster magnetic dynamics than ferromagnets (FM). Recently, honeycomb van der Waals (vdW) AFMs from the group $M$PS$_3$ ($M$ = Mn, Ni or Fe) have gained popularity for their layered structure, permitting tunable interlayer coupling, integration into heterostructures, and access to low-dimensional magnetism \cite{sivadas2015, takano2004, kim2019nps, kang2020, lee2016, chu2020, cui2023, bae2022}. Depending on the transition metal $M$, distinct AFM orders can emerge, giving rise to unconventional on-demand functionalities that are amenable to optical stimuli. Accordingly, experimental efforts using various time-resolved techniques have been made to uncover how ultrafast optical excitations disentangle lattice, charge, and magnetic degrees of freedom in these systems \cite{belvin2021, afanasiev2021nps, ergecen2022, zhou2022, ilyas2024, matthiesen2023}.

In MnPS$_3$, a honeycomb lattice in the $ab$ plane is formed by Mn atoms with alternating out-of-plane spins (Fig. \ref{fig:1}a), resulting in a N\'eel AFM order below $T_N=78$ K \cite{babuka2020,vaclavkova2020, wildes1998, sun2019}. A crucial feature that makes this material distinct from zigzag AFMs FePS$_3$ and NiPS$_3$ is that the AFM ordering in MnPS$_3$ breaks inversion symmetry which leads to non-zero second harmonic generation (SHG) below $T_N$ (Fig. \ref{fig:1}b--c) \cite{chu2020, ni2021, wang2024, muthukumar1995}. Essentially, the SHG in MnPS$_3$ is primarily linked with its spin texture since the structure remains centrosymmetric at all temperatures \cite{grasso1991, sun2019, kim2019mps}. Owing to this unique property, SHG combined with ultrafast techniques is well-suited for monitoring collective dynamics of magnetic order together with symmetry, structural distortion and electronic reconfiguration under strong-field laser excitation \cite{shan2021, luo2024, matthiesen2023}.

\begin{figure*}
    \centering
    \includegraphics[width=\linewidth]{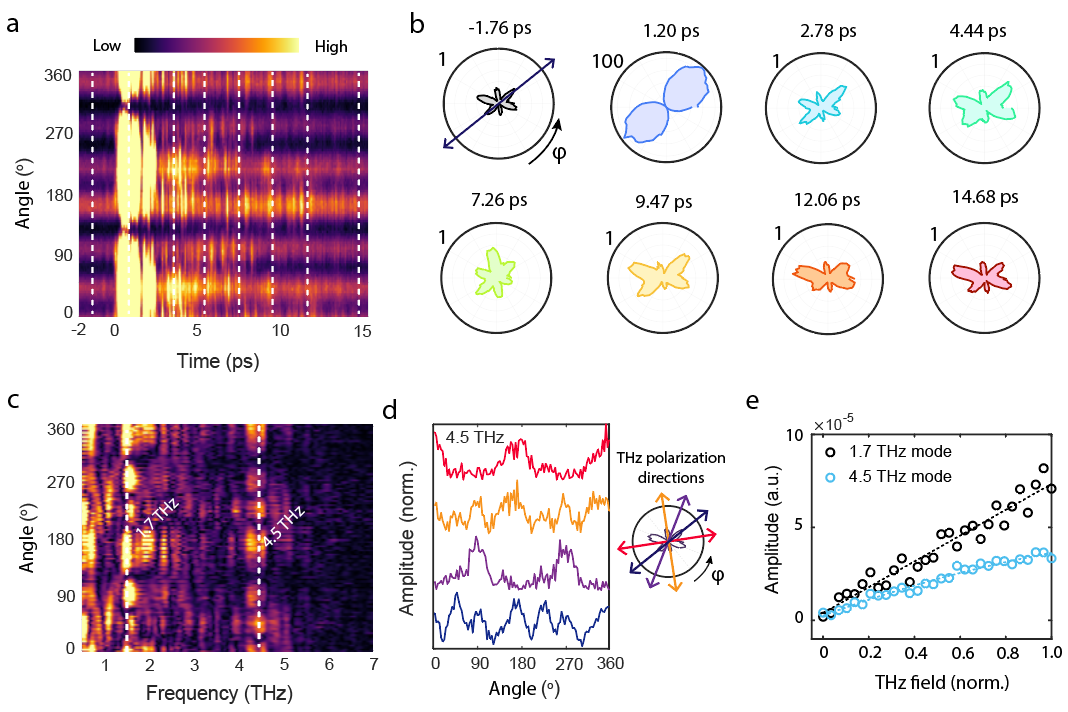}
    \caption{\textbf{Time- and angle-resolved SHG dynamics upon strong-field THz excitation. a}, Pump-induced SHG as a function of pump-probe delay time $t$ and polarizer analyzer angle $\varphi$ displayed in a normalized 2D intensity colormap at 9 K. Upon excitation with THz pulse with $\sim$ 500 kV/cm field strength at $t=0$, the TFISH response (0--2 ps) is followed by a long-lived oscillatory signal. A periodicity of $60\degree$ along the angular direction marks the 6 lobes of the SHG-RA pattern. \textbf{b}, SHG-RA patterns at different pump-probe delays, obtained from vertical linecuts in \textbf{a} (dashed white lines). The radial scale of each polar plot is normalized, with the outer circle corresponding to an amplitude of 1. The dark blue arrow indicates the THz pump polarization relative to the static SHG-RA pattern. For details, see the movies attached in the Supplementary Materials. \textbf{c}, Fourier transform spectrum of the pump-probe signal, revealing the coherent modes at 1.7 THz and 4.5 THz, particularly strong at lobe positions. \textbf{d}, Angular dependence of the 4.5 THz mode as a function of THz polarization. The colored arrows denote polarization directions relative to the SHG-RA pattern. \textbf{e}, Mode amplitudes as a function of normalized THz field, showing predominantly linear scaling. }
    \label{fig:2}
\end{figure*}

\subsection*{Terahertz-induced dynamics and giant oscillations}
\label{sec:SHG}
We use an 800 nm optical probe pulse, time-delayed with respect to a THz excitation (pump) pulse, to generate the SHG signal at 400 nm. By utilizing a time- and angle-resolved SHG system (Fig.~\ref{fig:1}d), we track the pump-induced dynamics in the SHG response of $\rm{MnPS_3}$ as a function of the pump–probe delay and the polarizer analyzer angle $\varphi$. Fig. \ref{fig:1}e presents the temporal evolution of the THz-induced SHG intensity $I_{SH}$, measured at a lobe position of the static SHG rotational anisotropy (SHG-RA) pattern (red circle), with the THz pump polarization (peak field $\sim$500 kV/cm) aligned along the same lobe direction (dark blue arrow). Data are shown for temperatures below (9 K, blue) and above (90 K, red) the N\'eel temperature $T_N$. The earlier-time dynamics (0--2 ps, yellow-shaded region) which is dominated by a sharp increase in SHG when the THz field is on, corresponds to THz field-induced second harmonic -- TFISH \cite{Cook1999,Chen2015}. This is an instantaneous response and is predominantly quadratic in THz field. For comparison, the temporal profile of the pump pulse is depicted with a vertical offset (black curve). We note that while the 90 K signal remains featureless at later times, the 9 K signal exhibits pronounced oscillations with amplitudes on the same order as the equilibrium SHG intensity (inset). The fast Fourier transform (FFT) of the signal in Fig. \ref{fig:1}f reveals two pronounced peaks: one at 1.7 THz, attributed to a theoretically predicted weakly Raman-active phonon with $B_g$ symmetry that has not been observed in equilibrium Raman spectroscopy \cite{babuka2020,vaclavkova2020}, and another at 4.5 THz, associated with modes that are both infrared- (IR) and Raman-active (symmetry: $A_{g/u}, B_{g/u}$), as seen in equilibrium spectra (Supplementary Fig. 1) \cite{vaclavkova2020,luo2024}. 
 
 The intensity colormap in Fig. \ref{fig:1}g plots the FFT spectra as a function of temperature. It is observed that as the temperature increases, the 1.7 THz and 4.5 THz mode amplitudes decrease, vanishing near $T_N$ (dashed horizontal line). Vertical linecuts highlight the order parameter-like temperature dependence of the 1.7 THz (Fig. \ref{fig:1}h) and 4.5 THz (Fig. \ref{fig:1}i) modes. These results unambiguously indicate that the observed giant coherent phonon signal emerges only in the AFM phase, underscoring the role of magnetically-induced inversion symmetry-breaking to enable coupling to the THz field.

\subsection*{Spectral fingerprints of symmetry modulation}
\label{sec:Results}
Figure \ref{fig:2}a presents the full time evolution of the SHG signal at 9 K as a function of analyzer angle, displayed as a 2D intensity colormap. In equilibrium (i.e., at negative time delays), a $60\degree$ periodicity is observed, consistent with the six-lobe structure of the static SHG-RA pattern (Fig. \ref{fig:1}b). As the THz pulse (polarized along a lobe of the SHG-RA pattern) arrives at $t=0 $ ps, the signal is initially dominated by the TFISH response. At later times, beyond the TFISH regime, clear long-lived coherent oscillations appear, persisting up to $\sim$15 ps. Vertical linecuts (dashed white lines) plot SHG-RA patterns at different pump-probe delays as illustrated in Fig. \ref{fig:2}b. Although six-lobe SHG-RA patterns reappear at later times ($t > 2.5$ ps), they show notable deviations from the nearly-symmetric equilibrium pattern observed at $t = -1.76$ ps (black polar plot), exhibiting unequal lobe intensities and distortions that break the original mirror symmetry (see the movies in the Supplementary Materials). 

Coherent modes at 1.7 THz and 4.5 THz are detected in the FFT spectra (Fig. \ref{fig:2}c). When the THz field is applied along a lobe of the SHG-RA pattern, the mode amplitudes exhibit a $60^\circ$ periodicity, mirroring the angular dependence shown in Fig.~\ref{fig:2}a. Strikingly, when the THz pulse is polarized along the nodes of the SHG-RA pattern, the 4.5 THz mode shows a $180^\circ$ periodicity outlined in Fig. \ref{fig:2}d (more details in Extended Data Figs. \ref{SHG_pol2}--\ref{SHG_pol4}), inferring a strong dependence of mode symmetry upon the pump polarization. As shown in Fig. \ref{fig:2}e, both 1.7 THz and 4.5 THz mode signals exhibit a notable linear dependence on the THz field strength, indicative of a linear excitation process (see also Extended Data Fig. \ref{field_dependence}).

\begin{figure*}
    \centering
    \includegraphics[width=\linewidth]{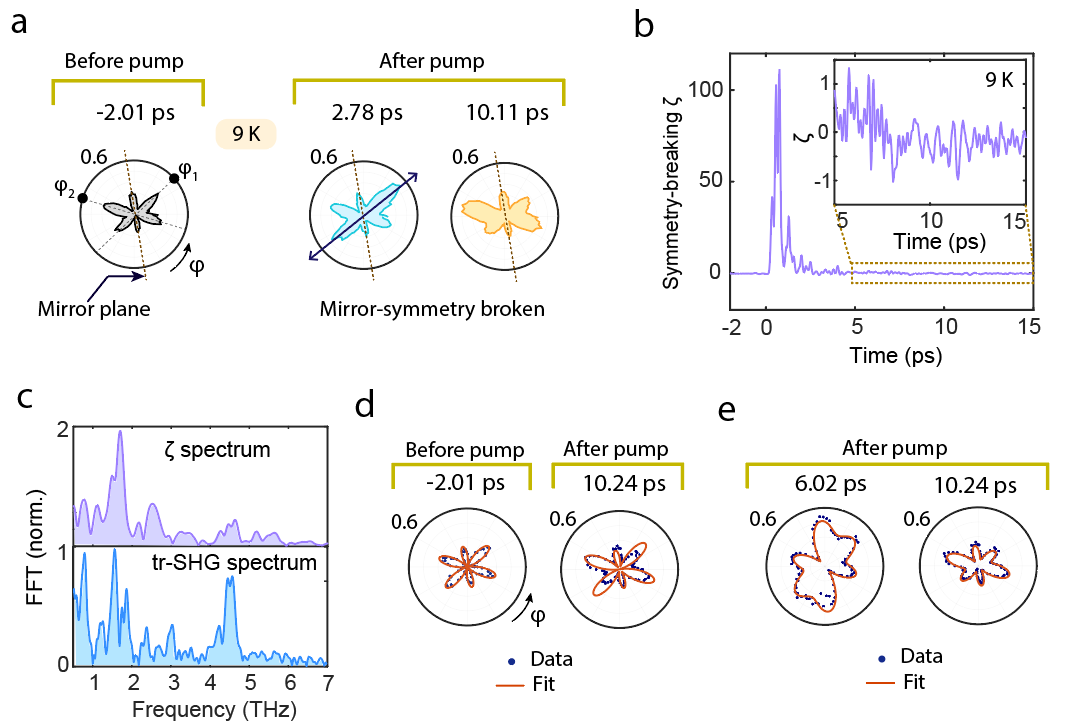}
    \caption{\textbf{THz field-induced long-lived mirror symmetry modulations. a}, SHG-RA patterns in equilibrium (black), and at different pump-probe time delays upon THz excitation (light blue and yellow) at 9 K. Photoexcited SHG-RA patterns exhibit a notable symmetry modulation relative to the equilibrium mirror plane (brown dashed line). The THz pump field is polarized along a lobe (dark blue arrow). \textbf{b}, Time-evolution of symmetry-breaking parameter $\zeta$, unmasking coherent oscillations at later times (inset). \textbf{c}, Fourier transform of $\zeta$ (top), compared with tr-SHG spectrum (bottom), demonstrating peaks at 1.7 THz and 4.5 THz. \textbf{d}, Equilibrium SHG-RA pattern at $t = -2.01$ ps fitted with complex parameters $A$ and $B$; this model does not reproduce the photoinduced patterns after THz excitation (i.e., $t = 10.24$ ps). \textbf{e}, Close agreement between nonequilibrium data and phenomenological fitting using complex parameters $A, B, C$ and $D$. The radial scale of each polar plot is normalized, with the outer circle corresponding to an amplitude of 0.6. }
    \label{fig:3}
\end{figure*}

Figure \ref{fig:3}a demonstrates that the equilibrium SHG-RA pattern measured at $t=-2.01$ ps is symmetric with respect to its mirror plane (brown dashed line corresponding to the $ac$-plane) \cite{ni2021}. In contrast, the nonequilibrium patterns at later times -- i.e., $t=2.78$ ps and $10.11$ ps -- show a clear breaking of mirror symmetry, manifested in unequal intensities between lobes connected by the mirror plane. To quantify the degree of mirror symmetry modulation in the SHG-RA patterns, we define a normalized asymmetry parameter $\zeta(t)=\frac{\int_{\varphi_1 - \Delta}^{\varphi_1 + \Delta} I_{SH}(t, \varphi) d\varphi-\int_{\varphi_2 - \Delta}^{\varphi_2 + \Delta} I_{SH}(t, \varphi)d\varphi}{\int_{\varphi_1 - \Delta}^{\varphi_1 + \Delta} I_{SH}^0( \varphi) d\varphi}$. Here, $I_{SH}(t, \varphi)$ is the time-resolved SHG (tr-SHG) intensity, $I_{SH}^0(\varphi)$ is the static SHG intensity, $\varphi_1$ and $\varphi_2$ are mirror-symmetric angles, and $\Delta$ defines the angular integration window. The dynamics of $\zeta$ at 9 K is displayed in Fig. \ref{fig:3}b, with oscillatory signal emerging at later times (inset). Fig. \ref{fig:3}c shows the FFT spectrum which shows the 1.7 THz and 4.5 THz modes (top), in accord with the tr-SHG spectrum (bottom). Overall, this striking resemblance in spectral fingerprints identifies the phonons as key contributors to the dynamical symmetry-breaking process.  

We examine the possibility of phenomenologically fitting the tr-SHG patterns as a function of analyzer angle $\varphi$ using the fitting function $I_{SH}=|A \cos^2(\varphi) \sin(\varphi) + B \sin^3(\varphi)|^2$ which corresponds to the expected functional form of the polar patterns based on the equilibrium symmetry \cite{chu2020}. Here, $A$ and $B$ are complex fit parameters, and depend on the elements of susceptibility tensor. Although the model accurately describes the equilibrium SHG-RA patterns, it falls short in capturing the symmetry-breaking features present in the nonequilibrium data (Fig. \ref{fig:3}d), necessitating the inclusion of additional terms. To address this, we introduce two new complex time-dependent parameters $C$ and $D$, which account for nonlinear susceptibility components that become symmetry-allowed in the reduced-symmetry nonequilibrium state. This leads to an extended fitting function: $I_{SH}=|A \cos^2(\varphi) \sin(\varphi) + B \sin^3(\varphi)+C \sin^2(\varphi) cos(\varphi) +D \cos^3(\varphi)|^2$, yielding a good agreement with the nonequilibrium SHG-RA patterns at later times as shown in Fig. \ref{fig:3}e (for details, see Supplementary Materials).

Importantly, the nonequilibrium parameters ($C$ and $D$) are comparable in magnitude to the equilibrium ones ($A$ and $B$), substantiating a strong and distinctive dynamical modulation of symmetry mediated by driven coherent phonons. A similar analysis for THz polarization along a node is presented in Supplementary Fig. 5.

\subsection*{Phonon-mediated magnetic manipulation via field-induced charge rearrangement}
\begin{figure*}[h!]
    \centering
    \includegraphics[width=\linewidth]{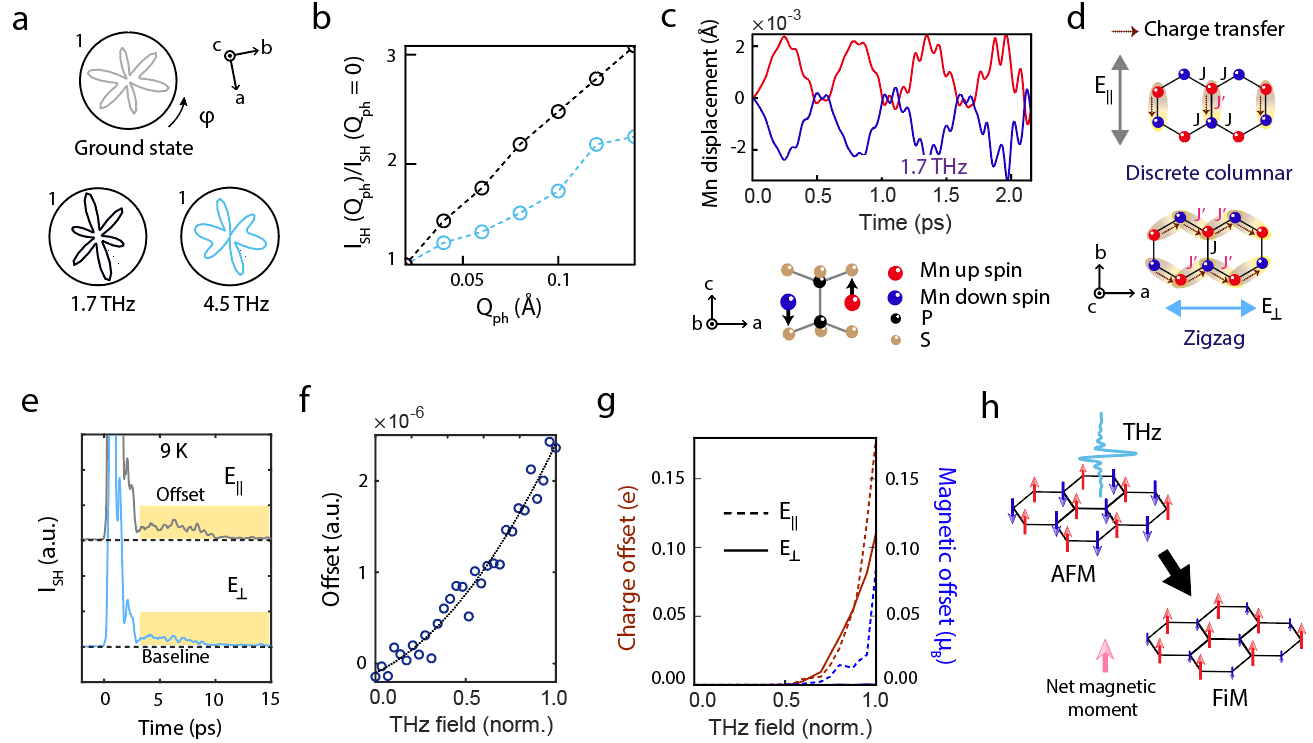}
    \caption{\textbf{Microscopic mechanism of phonon excitation. a}, SHG-RA patterns simulated via rt-TDDFT and frozen phonon displacements for low-frequency Raman-active modes, consistent with experimental observations. \textbf{b}, Computed relative change in SHG intensity scales linearly with the phonon amplitude $Q_{ph}$ for Raman modes. \textbf{c}, Simulated out-of-plane motion of opposite-spin Mn atoms (red and blue curves) matches the displacement pattern of the 1.7~THz $B_g$ phonon mode. \textbf{d}, THz-induced charge transfer modifies the magnetic exchange $J$ differently depending on THz polarization. For $E_{\parallel}$, charge is transferred between Mn sites with opposite spins in a discrete columnar pattern. $J'$ denotes the modified exchange parameter. For $E_{\perp}$, symmetric charge redistribution along Mn--Mn zigzag chains results in zero net charge transfer. Color gradient serves as a guide to the eye. \textbf{e}, A long-lived offset appears in the measured tr-SHG response. Black dashed line indicates equilibrium response and the offset is computed as the average difference from the baseline, evaluated after $\sim$4 ps. \textbf{f}, The offset shows nonlinear field dependence. \textbf{g}, Offset in charge transfer and magnetic moment obtained from simulations. For $E_\parallel$, a finite magnetic moment is observed, implying a transition to ferrimagnetic (FiM) state. \textbf{h}, Schematic of THz-induced potential AFM-FiM transition. Pink arrow indicates net magnetic moment.}
    \label{fig:4}
\end{figure*}

Insight into the SHG response of MnPS$_3$ can be gained using real-time time-dependent density functional theory (rt-TDDFT) simulations based on our proposed tight-binding model. The model includes hopping processes between Mn–-Mn, Mn–-S, and S–-S atomic pairs, involving Mn 3$d$ and S 3$p$ orbitals, as well as electronic interactions among Mn 3$d$ electrons (Extended Data Fig. \ref{DFT}a). The effect of the probe field $E_\omega$ (frequency $\omega$) is embedded into the model through light-matter coupling, using the Peierls substitution method \cite{Peierls1933}. The SHG yield is extracted from the peak at $2\omega$ in the Fourier transform of the current density computed using rt-TDDFT along the probe polarization direction (see Methods). Repeating this procedure across all polarization angles allows for the reconstruction of the static SHG-RA pattern, as displayed in Fig. \ref{fig:4}a (gray polar plot).

Introducing atomic displacements corresponding to low-frequency phonons into the equilibrium SHG structure can delineate how phonon modes influence the SHG-RA pattern and its mirror symmetry. Figure~\ref{fig:4}a shows the computed SHG-RA patterns with atomic displacements frozen along the 1.7 THz and 4.5 THz Raman-active modes. Each mode induces pronounced deviations from the ground state pattern, exhibiting distinct modulations of the SHG symmetry (see Methods and Extended Data Fig.~\ref{DFT}b--e). We note that the SHG-RA patterns show uneven mirror-connected lobes, in concert with our experimental observations of dynamical symmetry modulation, further supporting the role of select Raman-active phonons. The calculated SHG intensity $I_{SH}$ at a lobe as a function of phonon amplitude $Q_{ph}$ is shown in Fig.~\ref{fig:4}b, revealing a linear scaling for all Raman modes. At sufficiently large phonon amplitudes (i.e., 2\% of the equilibrium Mn--S bond length), the SHG intensity modulation closely matches that observed in experiment. 

Given the importance of phonons in the SHG symmetry, we examine the microscopic mechanism by which these modes are coherently driven by the THz field and exhibit linear field dependence. While there is a known IR-active 4.5 THz mode that can be resonantly excited by the tail of the THz pump pulse spectrum (Extended Data Table 2), the 1.7 THz mode is largely optically inactive, not observed in equilibrium spectroscopy (Supplementary Fig. 1), and therefore requires a unique field-driven interaction.  We can further rule out excitation via the THz magnetic field $B_{THz}$, since it only drives in-phase Mn spin precession without exciting any phonons or magnons, as confirmed by the rt-TDDFT calculations with external magnetic field (Extended Data Fig. \ref{DFT}f). Hence, the THz electric field $E_{THz}$ provides the viable excitation pathway for the phonon modes. 

$E_{THz}$ shifts the on-site energies of Mn $d$ and S $p$ orbitals, modifying Mn--S hybridization and enabling charge transfer between Mn and S atoms \cite{yamakawa2021}. This charge redistribution modulates the local magnetic moments at Mn sites and dynamically distorts the crystal field environment of the surrounding S atoms, in turn driving phonons through electron--phonon coupling. To investigate this possibility, we performed constrained DFT calculations in which 0.2\% of the electrons were transferred from the top of the valence band to the bottom of the conduction band, primarily populating minority-spin Mn $d$ orbitals. This setup captures essential features of the electronic  response induced by a THz pulse -- namely, hybridization between valence and conduction states, and a reduction in the magnitude of the local magnetic moments. Relaxing the atomic structure under this fixed electron distribution reveals distortions consistent with Raman-active modes, including those at 1.7 THz and 4.5 THz. At 4.5 THz frequency, IR-active modes can also be directly driven by the THz field. However, the eigenvectors of the 4.5 THz IR-active $A_u$ and $B_u$ modes feature pronounced displacements of P atoms, whereas the Raman-active $A_g$ and $B_g$ modes display minimal phosphorus involvement (Extended Data Fig \ref{4d5}a--b). In contrast, according to our calculations, the SHG response in MnPS$_3$ is primarily sensitive to the Mn--S bond current. This, together with the small THz pump spectral weight around 4.5 THz (see Extended Data Fig. \ref{SHGexeriment}), suggests that the 4.5 THz response in tr-SHG measurement contains comparable contributions from both the IR-active modes directly excited by the THz field and Raman-active modes excited via the field-induced charge transfer mechanism (see also Supplementary Fig. 12).

Having clarified the 4.5 THz feature, we next consider the more subtle 1.7 THz response. The constrained DFT calculations show that the THz-induced reduction of the local Mn magnetic moments is accompanied by out-of-plane displacements of the opposite-spin Mn atoms in opposite directions, as depicted in Fig.~\ref{fig:4}c. This motion is reminiscent of the the 1.7 THz eigendisplacement~\cite{babuka2020}, thereby suggesting a possible pathway to exciting this mode. Although multiple Raman- and IR-active modes are coherently driven under THz excitation, the 1.7 THz mode is particularly distinctive due to its absence in equilibrium spectra as well as in THz-induced birefringence/magneto-optic measurements (Supplementary Figs. 1 and 4). However, its appearance in tr-SHG underscores that its visibility under THz excitation requires sensitivity to the symmetry modulation at play. While the 1.7 THz mode is, in principle, weakly Raman-active, its linear field dependence also implies that a purely Raman-driven pathway alone cannot fully account for the experimental observations. Instead, its excitation may be better understood within the charge transfer framework, with weak IR activity -- enabled by inversion symmetry breaking due to AFM order and potentially enhanced by imperfections or defects -- offering an additional channel for coupling to the THz field \cite{cortijo2021,sander2014}.

Thus, our work reveals an excitation mechanism based on a symmetry-selective, field-driven process that enables tr-SHG to access vibrational modes otherwise inaccessible via conventional linear spectroscopic techniques. Moreover, the appearance of the 1.7 THz mode, which is sensitive to change in magnetic moment, serves as a signature of dynamical magnetic manipulation by the THz field. 

Additional support for the charge rearrangement process is found in the THz polarization dependence as showcased in Fig. \ref{fig:4}d. In the AFM phase, when the THz field is polarized along  nearest-neighbor Mn sites ($E_{\parallel}$), charge is transferred between Mn atoms with opposite spins in a discrete pattern. In comparison, for the THz field polarized perpendicular to Mn--Mn bonds ($E_\perp$), the charge rearrangement occurs along Mn--Mn zigzag chains and contributions from neighboring Mn sites cancel out, resulting in zero net charge transfer. Therefore, distinct THz-induced dynamics may arise depending on the field polarization direction. In fact, this is exactly what is observed: a six-fold symmetry of the 4.5 THz mode appears in the tr-SHG spectrum when the THz field is polarized along a lobe of the SHG-RA pattern (analogous to $E_{\perp}$), and a two-fold symmetry emerges when the field is aligned along a node ($E_\parallel$), as shown in Fig. \ref{fig:2}d. This behavior can be understood as a polarization-dependent preferential excitation of 4.5~THz IR- and Raman-active modes with different symmetries -- $A_{g/u}$ (excited by $E_{\parallel}$) and $B_{g/u}$ (excited by $E_{\perp}$) -- accompanied by the anisotropic renormalization of specific Mn--S bonds. With $E_\parallel$ applied, the bonding strength and hybridization of the six Mn--S bonds surrounding each Mn atom are altered along a single Mn--S bond direction giving rise to a two-fold response; in contrast, for $E_\perp$, this is modulated along all three Mn--S directions, resulting in a six-fold response (more details in Methods and Extended Data Fig. \ref{4d5}). The IR modes ($A_u$, $B_u$) are directly excited by the THz field whereas the charge rearrangement process drives the Raman modes ($A_g$, $B_g$). Since the $A_{g}$ and $B_{g}$ modes are known to be sensitive to AFM ordering \cite{sun2019,vaclavkova2020}, we argue that the observed dynamics arise from polarization-selective modulation of the AFM exchange interaction, activated via charge redistribution pathways. 

\subsection*{Toward a photoinduced magnetic phase}
Simulations corroborate several key features of this mechanism. As presented in Extended Data Fig. \ref{DFT}g, notable charge transfer from S to Mn atoms leads to an increased charge density at Mn sites. Temporal modulation of the local Mn magnetic moments was also observed (Extended Data Fig. \ref{DFT}h). Moreover, the degree of charge transfer from S to Mn atoms increases monotonically with pump field strength in the low-field regime (Extended Data Fig. \ref{DFT}i--j), compatible with an excitation process that is linear in field as observed experimentally. In particular, at higher fields, the simulations predict a long-lived offset (rectification) in the response (insets of Extended Data Fig. \ref{DFT}g--h), consistent with the experiments (Fig. \ref{fig:4}e). The magnitude of the offset is strongly polarization-dependent, reaching higher values for $E_\parallel$ (gray curve in Fig. \ref{fig:4}e) than for $E_\perp$ (blue curve). As sketched in Fig. \ref{fig:4}f, the offset grows nonlinearly with the pump field. Likewise, Fig. \ref{fig:4}g shows simulated long-lived offsets in charge transfer from S to Mn (red curves) and net magnetic moment (blue curves), both exhibiting nonlinear dependence on the THz field. Dashed and solid lines correspond to $E_{\parallel}$ and $E_{\perp}$ polarizations, respectively. Remarkably, a net magnetic moment on the Mn atoms emerges only for $E_{\parallel}$ (dashed blue line), signaling a field-induced tendency towards ferrimagnetic (FiM) order (Extended Data Fig. \ref{4d5}c--d). Appearance of this net moment for $E_\parallel$ may also explain the enhanced offset measured in Fig.~\ref{fig:4}e, pointing to a correspondence between the simulated magnetic transition and the polarization-sensitive SHG signal. Simulations further predict that the net magnetic moment increases with field strength which suggests a possible transition to ferromagnetic order at sufficiently high fields. Together, these results establish the pivotal role of charge rearrangement in THz-induced magnetic transitions, linking field-dependent charge and spin dynamics to dynamical symmetry modulation in MnPS$_3$. Further experiments under controlled magnetic fields will be an important future step to conclusively establish the FiM phase. 

\subsection*{Outlook}

Our results demonstrate new mechanisms by which  atomic and magnetic degrees of freedom in vdW antiferromagnetics can be dynamically controlled by THz fields. Time-resolved SHG measurements show that specific Raman-active phonons are efficiently excited via linear-in-field mechanisms linked to charge redistribution. This modifies the nonlinear susceptibility in a manner that breaks mirror symmetry and launch large-amplitude oscillations in the SHG-RA patterns. The polarization dependence of the symmetry modulation dynamics, the observation of new phonon modes (specifically at 1.7 THz), and correlated first principles theory provides support for this and points towards the possibility of accessing new metastable magnetic phases. Extending this framework to other quantum magnets or inversion symmetry-broken systems may enable targeted control of phase transitions, coherent switching between magnetic states, or the design of advanced nonlinear optical functionalities. More broadly, our findings provide a roadmap for strong-field vibrational control as a versatile tool for tailoring symmetry and magnetism in quantum materials.

\newpage
\section*{Methods}
\label{meth}
\subsection*{Sample preparation}
For the time-resolved measurements, 4 mm × 3 mm MnPS$_3$ single crystals with $\sim$100 $\mu m$ thickness were used (2D Semiconductors). For the infrared measurements, millimeter-scale single crystals were grown using a chemical vapor transport (CVT) method. A mixture of high-purity elements, namely manganese (Mn), phosphorus (P), and sulfur (S) in the ratio of 1:1:3.03, along with 0.05 g of iodine as the transport agent was sealed in an evacuated quartz ampoule. The tube was then placed in a horizontal two-zone furnace.  Initially, the temperature was raised up to 400°C within 6 hours at the source zone  (containing the reactants ) and to 450°C in the growth zone and held at these temperatures for an additional 6 hours. Subsequently, the temperature was increased up to 650°C at the source zone and to 600°C in the growth zone within 3 hours. After a week, the tube was removed from the furnace and allowed to cool in air. The resulting crystals, measuring between 1 and 10 mm, were washed in isopropanol to remove the halogen from their surface.

\subsection*{THz pump -- time-resolved SHG probe experiment}
Extended Data Fig. \ref{SHGexeriment}a shows a schematic of our experiment. We employed a Coherent Legend Elite regenerative amplified Ti:sapphire laser system generating pulses with 3–5 mJ energy, 70 fs duration, and 800 nm central wavelength at a repetition rate of 700 Hz. The output beam was split into two paths using a beam splitter: one arm was directed towards generation of the terahertz (THz) pump pulses, while the other was used to produce probe pulses. 

The pump arm of the beam passed through a custom-made multi-pass amplifier, yielding 16 mJ pulses at 800 nm with a reduced repetition rate of 350 Hz. These pulses were coupled into a TOPAS-TWINS-HE system (Light Conversion), composed of two tunable 2-stage optical parametric amplifiers (OPA).  For THz generation, we used the signal output of one OPA tuned to 1.3 $\rm \mu m$, delivering 2.5 mJ pulses ($\sim$ 70 fs duration). This beam was used to pump a 500 $\rm \mu m$-thick DSTMS crystal (4-N,N-dimethylamino-4’-N’-methyl-stilbazolium 2,4,6-trimethylbenzenesulfonate) in a nitrogen-purged environment to generate broadband THz radiation via optical rectification, having peak field amplitude of ~500 kV/cm and beam spot diameter of ~380 $\rm \mu m$ (FWHM). Note that the generation crystal is on a rotatable mount and a half-wave plate is placed in the 1.3 $\rm \mu m$ beam path, allowing for THz polarization control. The THz beam was then focused onto the sample inside a vacuum chamber using off-axis parabolic mirrors in a normal incidence geometry. Additional measurements were also performed using an OH1 generation crystal (2-[3-(4- hydroxystyryl)-5, 5-dimethylcyclohex-2-enylidene] malononitrile) for comparison. In order to characterize the THz pump, the time-domain THz signals in vacuum were acquired using electro-optic sampling (EOS) method. The EOS signal as well as the corresponding FFT for DSTMS (OH1) cystal are displayed in Extended Data Fig. \ref{SHGexeriment}b and \ref{SHGexeriment}d (\ref{SHGexeriment}c and \ref{SHGexeriment}e), respectively.

For the SHG probe, the 800 nm pulse was passed through a zero-order quartz half-wave plate (HWP) for polarization control and was focused normally onto the sample. The transmitted beam contained both the fundamental and the second harmonic (400 nm). To isolate the SHG signal, we used a narrowband filter stack (FGB39M + FBH-400-40) to block the fundamental, followed by a rotating Glan-Thompson polarizer acting as an analyzer. This analyzer was rotated at twice the frequency of the HWP to enable full 360° SHG-RA mapping. 

The SHG signal was detected using a photomultiplier tube (Hamamatsu H10722-110) using a Stanford Research Systems lock-in amplifier (Model SR830) which was triggered by a 700 Hz TTL reference signal. We limited the probe energy to 1.3 $\rm \mu J$ to prevent sample heating or damage. Throughout this work, only the SHG signal parallel to the 800 nm probe polarization was considered since the cross-polarized SHG yield was several orders of magnitude weaker.

\subsection*{THz field-induced SHG signal and phonon modulation}
For an inversion symmetry-broken crystal, the second-order polarization response to an input light (probe) with frequency $\omega$ can be written as $P(2\omega)=\chi^{(2)}E_\omega^2$. Here, $\chi^{(2)}$ and $E_\omega $ are the second-order nonlinear susceptibility and probe field, respectively. The corresponding equilibrium SHG intensity will be $ I_{SH}^0\propto |P(2\omega)|^2=|\chi^{(2)}E_\omega^2|^2$. 

Now, we consider the material to be subjected to a THz pump excitation. As the THz pump is incident upon the sample, the initial dynamics (first few picoseconds) is dominated by a sharp peak and an incoherent background (Fig. \ref{fig:1}e). This response is termed as THz field-induced second harmonic (TFISH), and is attributed to instantaneous inversion symmetry-breaking by the THz field $E_{THz}$. As a result, the renormalized second-order polarization is given by $P(2\omega)=\chi^{(2)}E_\omega^2+\chi^{(3)}E_\omega^2 E_{THz}$, where $\chi^{(3)}$ is the third-order nonlinear susceptibility. The SHG intensity is
\begin{equation}
    I_{SH}\propto |P(2\omega)|^2=|\chi^{(2)}E_\omega^2+\chi^{(3)}E_\omega^2 E_{THz}|^2.
    \label{total_SHG_pump}
\end{equation}

From these relations, pump-induced change in SHG intensity is expressed as:
\begin{align}
\Delta I_{SH} &=I_{SH}-I_{SH}^0 \propto|\chi^{(2)}E_\omega^2+\chi^{(3)}E_\omega^2 E_{THz}|^2-|\chi^{(2)}E_\omega^2|^2 \\
&=\chi^{(3)2}E_\omega ^4 E_{THz}^2+2 \chi^{(2)}\chi^{(3)}E_\omega^4 E_{THz}.
\label{SHG_change}
\end{align}

From Eq. (\ref{SHG_change}), it appears that $\Delta I_{SH}$ response contains terms both linear and quadratic in $E_{THz}$. At earlier times, while $E_{THz}$ is present, the pump-probe signal is dominated by the first term in Eq. (\ref{SHG_change}) \cite{luo2024}. This term accounts for the TFISH response characterized by a quadratic dependence in $E_{THz}$. However, at later times where $E_{THz}\sim0$, the dynamics is dominated by phonon modes, manifesting in long-lived coherent oscillations. Along this line, $E_{THz}$ is replaced by phonon coordinate $Q_{ph}$ leading to the pump-probe signal
\begin{equation}
    \Delta I_{SH}(t)=\chi^{(3)2}E_\omega ^4 Q_{ph}^2(t)+2 \chi^{(2)}\chi^{(3)}E_\omega^4 Q_{ph}(t).
    \label{coherent_SHG}
\end{equation}

Here, $\chi^{(3)} Q_{ph}$ denote the phonon-induced change in static second-order susceptibility and $\chi^{(3)} Q_{ph}\ll\chi^{(2)}$ for a material with very large $\chi^{(2)}$. Thus, the first term in Eq. (\ref{coherent_SHG}) can be dropped and we argue that the second term will be the dominant contribution in the detected signal.

\subsection*{Theoretical formulation: multiband tight-binding model}
We construct a multiband tight-binding model for bulk MnPS$_3$, where ten degenerate Mn 3$d$ spin orbitals and six degenerate S 3$p$ orbitals are taken into account. The electrons are described by a set of one-particle wave functions. To capture noncollinear magnetic structures, each wave function is treated as a two-component spinor. The one-particle wave function with band index $n$ is expressed as
\begin{equation}
|\psi_n\rangle = \sum_{\sigma, \alpha, R} \chi_{\sigma, \alpha, R} \, \psi_{\sigma, \alpha, R, n}.
\label{spinor}
\end{equation}

In Eq. (\ref{spinor}), $\chi_{\sigma, \alpha, R}$ are local spin orbitals characterized by spin $\sigma = \{\uparrow, \downarrow\}$ and orbital index $\alpha \in \{p_x\!, p_y\!, p_z\!, d_{xy}\!, d_{yz}\!, d_{xz}\!, d_{x^2\! -\! y^2}\!, d_{3z^2\! -\! r^2}\}$. The $d$ orbitals are centered on Mn atoms and the $p$ orbitals on the surrounding S atoms. The total energy $E_{tot}$ of the model is given by 
\begin{equation}
    E_{tot}=E_{p}+E_{d}+E_{pd},
    \label{E_tot}
\end{equation}
where $E_{p}$ corresponds to bare atomic energies of the S $p$ orbitals
\begin{equation}
E_{p}=\epsilon_p\sum_{R',\alpha,\sigma}n_{\sigma,\alpha,R'}. 
\label{Ep}
\end{equation}

Notably, $E_{d}$ consists of the bare atomic energies of $d$ electrons at Mn sites and $d-d$ Coulomb interaction contribution $E_{e-e}$. This is expressed by
\begin{eqnarray}
E_{d}=\epsilon_d\sum_{R,\alpha,\sigma}(n_{\sigma,\alpha,R} + E_{e-e,R}).
\label{Ed}
\end{eqnarray}

Here, $n_{\sigma,\alpha,R}$ are the diagonal matrix elements of the local reduced density matrix centered at $R$-th Mn site which is defined as
\begin{eqnarray}
\rho_{\sigma_1\alpha_1,\sigma_2\alpha_2,R}=\sum_n f_n\psi^*_{\sigma_1, \alpha_1, R, n} \, \, \psi_{\sigma_2, \alpha_2, R, n},
\label{diagonal}
\end{eqnarray}
where $f_n$ is the occupation of the wavefunction $\psi_{\sigma_1, \alpha_1, R, n}$. Now, we consider on-site $d-d$ Coulomb interaction on the Hartree-Fock level \cite{Rajpurohit2024b,Sotoudeh2017} having the following form:
\begin{align}
E_{e\text{-}e,R} &= \frac{U}{2} \sum_{\substack{\sigma_1 \ne \sigma_2 \\ \alpha}} 
\Big( n_{\sigma_1\alpha,R}n_{\sigma_2\alpha,R} 
+ \rho_{\sigma_1\alpha,\sigma_2\alpha,R}\rho_{\sigma_2\alpha,\sigma_1\alpha,R} \Big) \nonumber \\
& + \frac{U{-}3J_{xc}}{2} \sum_{\substack{\sigma_1 \\ \alpha_1 \ne \alpha_2}} 
\Big( n_{\sigma_1\alpha_1,R}n_{\sigma_1\alpha_2,R} 
- \rho_{\sigma_1\alpha_1,\sigma_1\alpha_2,R}\rho_{\sigma_1\alpha_2,\sigma_1\alpha_1,R} \Big) \nonumber \\
& + \frac{U{-}2J_{xc}}{2} \sum_{\substack{\sigma_1 \ne \sigma_2 \\ \alpha_1 \ne \alpha_2}} 
\Big( n_{\sigma_1\alpha_1,R}n_{\sigma_2\alpha_2,R} 
- \rho_{\sigma_1\alpha_1,\sigma_2\alpha_2,R}\rho_{\sigma_2\alpha_2,\sigma_1\alpha_1,R} \Big) \nonumber \\
& - \frac{J_{xc}}{2} \sum_{\substack{\sigma_1 \ne \sigma_2 \\ \alpha_1 \ne \alpha_2}} 
\Big( \rho_{\sigma_1\alpha_1,\sigma_2\alpha_1,R} \rho_{\sigma_2\alpha_2,\sigma_1\alpha_2,R}
- \rho_{\sigma_1\alpha_1,\sigma_1\alpha_2,R} \rho_{\sigma_2\alpha_2,\sigma_2\alpha_1,R} \Big) \nonumber \\
& + \frac{J_{xc}}{2} \sum_{\substack{\sigma_1 \ne \sigma_2 \\ \alpha_1 \ne \alpha_2}} 
\Big( \rho_{\sigma_1\alpha_1,\sigma_1\alpha_2,R} \rho_{\sigma_2\alpha_1,\sigma_2\alpha_2,R}
- \rho_{\sigma_1\alpha_1,\sigma_1\alpha_2,R} \rho_{\sigma_2\alpha_1,\sigma_2\alpha_2,R} \Big).
\label{hartree_fock}
\end{align}

Here, $U$ and $J_{xc}$ denote Coulomb parameters. This expression contains terms similar to the local repulsion terms in Kanamori Hamiltonian \cite{Kanamori1963,Georges2013}. The first three terms, with prefactors $U/2$, $(U - 3J_{xc})/2$, and $(U - 2J_{xc})/2$, correspond to interactions between electrons with opposite spins in the same orbital, parallel spins in different orbitals, and opposite spins in different orbitals, respectively. The last two terms correspond to pair-hopping and spin-flip interactions. This expression for $E_{e\text{-}e,R}$ is rotationally invariant in both the orbital and spin spaces of the Mn $d$-electrons.

The energy $E_{pd}$ represents the kinetic contribution from hopping processes, and is given by
\begin{equation}
E_{pd} = \sum_{\langle R, R' \rangle} \sum_n f_n \sum_{\sigma} \sum_{\alpha, \beta} \psi^*_{\sigma, \alpha, R, n} \, t^{hop}_{\alpha, \beta, R, R'} \, \psi_{\sigma, \beta, R', n},
\label{E-pd}
\end{equation}
where $R$ and $R'$ denote the Mn or S site while $\alpha$ and $\beta$ represent the $d$ or $p$ orbital indices. The hopping contribution are evaluated for all atomic pairs with interatomic distances up to 4.00 \AA{}. The hopping $t^{hop}_{\alpha, \beta, R, R'}$ between sites are parameterized using directional-dependent Slater-Koster two-center integrals: $V_{pd\sigma }$, $V_{pd\pi}$,$V_{dd\sigma }$, $V_{dd\pi}$, $V_{pp\sigma }$, $V_{pp\pi}$. These parameters correspond to the $\sigma$- and $\pi$-bonding overlaps for the Mn--S, Mn--Mn, and S--S bonds. In the subscripts, $p$, $d$ represent the $p$, $d$ orbitals on the Mn atom and six surrounding S sites, respectively. 
Following Harrison’s approach, we introduce a distance dependence for the hopping amplitudes: \( t^{\text{hop}}_{pd} \propto V_{pd}/d^{4} \), \( t^{\text{hop}}_{dd} \propto V_{dd}/d^{5} \), and \( t^{\text{hop}}_{pp} \propto V_{pp}/d^{3} \). The direction-cosine-dependent Slater–Koster two-center integrals \( V_{\mu\nu} \) are computed using atomic positions obtained from DFT. The hopping terms are then expressed as \( t^{\text{hop}}_{pd} = A_{pd} V_{pd}/d^4 \), \( t^{\text{hop}}_{dd} = A_{dd} V_{dd}/d^5 \), and \( t^{\text{hop}}_{pp} = A_{pp} V_{pp}/d^3 \), where \( A \) is a proportionality constant. We assume \( A_{pd\sigma} = -A_{dd\sigma} = A_{pp\sigma} \), \( A_{pd\pi} = A_{pp\pi} = -A_{pp\sigma}/2 \), and \( A_{dd\pi} = -A_{dd\sigma}/4 \).

To extract the model parameters, we first determine the on-site energy levels of the Mn $d$ and S $p$ states in terms of the model parameters \(\epsilon_p\), \(\epsilon_d\), \(U\), and \(J_{xc}\) from the \(E_p\) and \(E_d\) parts of the Hamiltonian. Next, we consider: (1) the energy difference \(\Delta^{o}_{pd} = \epsilon_p - \epsilon_d + 4U - 12J_{xc}\); and (2) the energy difference between the majority and minority Mn \textit{d} spin states, \(\Delta^{o}_{d\uparrow-\downarrow}=U+4J_{xc}\). A comparison between the model-predicted values of \(\Delta^{o}_{pd}\) and \(\Delta^{o}_{d\uparrow-\downarrow}\), and the corresponding quantities extracted from the DFT-calculated density of states (DOS) — namely, \((\epsilon^o_{p\uparrow} + \epsilon^o_{p\downarrow})/2 - (\epsilon^o_{d\uparrow} + \epsilon^o_{d\downarrow})/2\) and \(\epsilon^o_{d\uparrow} - \epsilon^o_{d\downarrow}\) — provides conditions to determine the values of the model parameters. 
Here $\epsilon^o_{p\uparrow/\downarrow}$ and $\epsilon^o_{d\uparrow/\downarrow}$ are estimated from the first moments (centers of mass) of the spin-resolved projected DOS. Furthermore, we consider $J_{xc}=U/3$. 
The comparison of the energy difference $\Delta E_{FM-AFM}$ between the AFM and FM states predicted by the full model  with the corresponding DFT values yields the hopping constant $A_{\mu\nu}$. Details of the model parameters are given in Extended Data Table \ref{tab:t1}.

\subsection*{\textit{Ab-initio} electronic structure calculations}
We performed density functional theory (DFT) calculations for bulk MnPS$_3$ using the projector augmented wave (PAW) method \cite{Bloechl1994}. A 1×2×1 supercell of the $P21/n$ space group, containing 40 atoms (eight formula units) and the high-temperature experimental lattice constants, $a=5.935$ {\AA}, $b=5.935$ {\AA}, and $c=6.596$ {\AA} are used. The calculations employed a 4×2×4 $k$-point grid and a plane-wave cut-off energy of 35 $Ry$. The DFT calculations were carried out using the PBE0r hybrid functional used previously for studying other $3d$ transition-metal compounds \cite{Sotoudeh2017, Eckhoff2020, rajpurohit2024a}. The PBE0r functional is a range-separated extension of the global hybrid PBE0 functional \cite{Perdew1996} where the Hartree-Fock exchange is calculated by restricting the evaluation to on-site exchange contributions within a local orbital basis set. To obtain mixing factors for on-site exchange, only on-site exchange terms are included, while off-site and long-range exchange interactions are neglected in PBE0r. We choose $a_x=10$\% mixing factor for on-site exchange for Mn, S, and P atoms. Our calculation predicts band-gap of 2.36 eV.  

The results are illustrated in Extended Data Fig. \ref{DFT}a where we plot the total DOS of bulk MnPS$_3$ (gray) as well as its projections onto Mn $d$ (blue), S $p$ (yellow) and P $p$ orbitals (red). The Fermi level ($E=0$) is marked by the solid vertical line while solid and transparent colors designate occupied and unoccupied states, respectively. From the figure, it is noted that Mn $d$ and S $p$ are the dominant states near the Fermi level, implying that SHG from MnPS$_3$ is primarily sensitive to Mn-S hybridization. 

\subsection*{\textit{Ab-initio} phonon calculations}

Phonon calculations were carried out using density functional perturbation theory (DFPT) as implemented in QUANTUM ESPRESSO package \cite{giannozzi2017, giannozzi2009}. Calculations were performed in the primitive cell with $a=b=6.07$ {\AA} and $c=6.76$ {\AA}. The spin-polarized Kohn-Sham equations were solved on $2\times2\times2$ $k$-point mesh with a 80 $Ry$ plane-wave cutoff. We used an LDA exchange-correlation functional with norm-conserving pseudopotentials sourced from Pseudo-Dojo and applied an on-site Hubbard $U=2.5$ eV to Mn 3$d$ orbitals. 

Mode symmetries and IR activities are summarized in Extended Data Table~\ref{tab:phonon-calc} for phonons 4–11. Below $T_N$, AFM ordering reduces the crystal symmetry, and the modes are classified by how they transform under the reduced $C_2$ symmetry ($A$, $B$). Above $T_N$, the modes are classified according to the irreducible representations of the parent $C_{2h}$ point group ($A_{u}$, $A_{g}$, $B_{u}$, $B_{g}$). While our calculations assume AFM order and thus use the $A/B$ classification, we also include the parent symmetry labels, since they are useful for deriving approximate selection rules.

The calculated eigenvectors and frequencies for phonon modes 4-11 from our calculations are broadly consistent with earlier theoretical studies \cite{vaclavkova2020}, in addition to the small variations arising from the sensitivity of these modes to the choice of computational parameters. Modes 4, 10, and 11 are particularly relevant, as their frequencies closely match the 1.7 THz and 4.5 THz coherent oscillations observed in time-resolved SHG. The 4.5 THz modes (10 and 11) exhibit measurable IR activity, suggesting that they can be directly excited by the THz pump. Mode 10 ($A_u$) couples to $E_{\parallel}$ polarization, whereas mode 11 ($B_u$) couples to $E_{\perp}$. In contrast, the 1.7 THz mode has negligible IR activity. Above $T_N$, its $B_g$ symmetry forbids IR coupling; although AFM order formally relaxes this restriction below $T_N$, our calculations show that the coupling remains very weak. We therefore rule out direct excitation of this mode by the THz pump.

Next, we examine the efficiency with which this mode can couple to light through  second-order Raman processes. For $B_g$ modes, symmetry ensures that the $(E_\perp, E_\perp)$ and $(E_\parallel, E_\parallel)$ components are zero. However, the $(E_\perp, E_\parallel)$ components may be finite. To estimate the efficiency of this coupling, we compute the Raman tensor using finite difference: $R^\nu_{ij} \propto \frac{\partial \epsilon_{ij}}{\partial Q_\nu}$, where $\varepsilon_{ij}$ is the optical dielectric tensor and the derivative is with respect to freezing in an atomic displacement of mode $\nu$ with amplitude $Q_\nu$. Using this approach, we find that $(E_\perp, E_\parallel)$ component of the Raman tensor for mode 8 is 20.7 times larger than that of mode 4. This suggests that the Raman coupling for the 1.7 THz mode is intrinsically weak, thereby providing a plausible explanation for why it has not been observed in equilibrium experiments to date.

\subsection*{THz field-induced magnetic and electronic response}
To investigate the pump-induced electronic and magnetic response of the bulk MnPS$_3$, we simulate the charge dynamics under a THz excitation using the real-time time-dependent density functional theory (rt-TDDFT) framework based on the above proposed tight-binding model. The one-particle wavefunctions of electrons evolve according to time-dependent Schr\"odinger equation
\begin{equation}
i\hbar\frac{\partial\psi_{\sigma,\alpha,R,n}}{\partial t}=\frac{\partial E_{tot}}{\partial\psi^*_{\sigma,\alpha,R,n}},
\label{tdse}
\end{equation}
where $E_{tot}$ is defined in Eq. (\ref{E_tot}). The effect of the pump electric field $E_{THz}$ is incorporated into the hopping term via Peierls substitution \cite{Peierls1933}. Further, we introduce the THz magnetic field $B_{THz}$ in the model via an additional term accounting for Zeeman coupling at Mn sites $R_{Mn}$
\begin{equation}
E_{mag}= g \mu_{B} B_{THz}\sum_{\alpha,R\in R_{Mn}} (n_{\uparrow,\alpha,R}-n_{\downarrow,\alpha,R}).
\label{E_mag}
\end{equation}

Here, $n_{\uparrow,\alpha,R}$ is defined in Eqs. (\ref{Ed}) and (\ref{diagonal}), while $g$ and $\mu_{B}$ represent the coupling constant and Bohr magneton, respectively. Simulations predict that the magnetic field component of a THz pulse with an electric field amplitude varying from 0 to 15 MV/cm drives only in-phase precession of Mn spins, without generating magnon excitations. Extended Data Fig. \ref{DFT}f shows the evolution of the spin components $S_x$, $S_y$, and $S_z$ for the two Mn sites with opposite spin orientations, highlighting spin precession about the $z$-axis under a THz magnetic field applied along the $z$-direction. The spin vectors for Mn spin-up and spin-down sites in the ground state at $t=0$ are $\vec{S_{\uparrow}}=(-3.64, -2.82, 0.00)$ and $\vec{S_{\downarrow}}=(3.64, 2.82, 0.00)$, respectively.

The driving frequency is fixed at 2.5 THz in our simulations, while the peak-electric-field strength $|E_{THz}|$ is varied from 0 to 15 MV/cm. The electric field polarization is in the $ab$ plane, and is oriented either (1) parallel to the line connecting first nearest-neighbor Mn sites ($E_{THz} \parallel$ Mn--Mn bonds, denoted as $E_{\parallel}$), or (2) perpendicular to it, along the line connecting second nearest-neighbor Mn sites (denoted as $E_{\perp}$). The simulation cell contains 2 Mn atoms and 6 S atoms. We use  4×4×4 $\Gamma$-centered $k$-grid. Extended Data Fig. \ref{DFT}g show the evolution of local charge density at Mn sites, while Extended Data Fig. \ref{DFT}h displays the corresponding changes in the magnetic moment at Mn sites as a function of field strength.

In both the polarization cases ($E_{\parallel}$ and $E_{\perp}$), our simulations reveal charge transfer between Mn and S atoms during the presence of the pump field (Extended Data Fig. \ref{DFT}i). This transfer is driven by field-induced shifts in the on-site energy levels of Mn $d$ and S $p$ orbitals, which enhance the hybridization between majority-spin Mn $d$ and S $p$ states. As a result, the local magnetic moments at Mn sites are modulated, and the magnitude of the charge rearrangement scales linearly at lower field strengths (Extended Data Fig. \ref{DFT}j).

When a THz field is applied, the hybridization along the field polarization direction is enhanced, realizing in an anisotropic modulation in effective AFM exchange $J$, where $J \propto |t^2|/\Delta$, between Mn atoms. Here, $\Delta\approx (U-3J_{xc})$ denotes the on-site spin-splitting at Mn sites. In contrast, the interaction $J$ between Mn pairs whose bond directions are not aligned with the field polarization remains largely unchanged. The field with polarization parallel to Mn--Mn bonds ($E_{\parallel}$) modifies $J$ for one-third of the Mn--Mn pairs, while $E_{\perp}$ affects $J$ for two-thirds of the Mn--Mn pairs. For ${E}_{\perp}$, the induced charge transfer occurs along the Mn--Mn zigzag chains. However, at any given Mn site, the contributions from neighboring Mn atoms on either side are symmetric and cancel each other out. Consequently, there is no net charge transfer between Mn atoms. However, for ${E}_{\parallel}$, the Mn--Mn connectivity is discontinuous, manifesting an imbalance and a net charge transfer between Mn sites (Fig. \ref{fig:4}d). 

The observed linear field dependence of both the average reduction in magnetic moment and the S-to-Mn charge transfer (Extended Data Fig. \ref{DFT}h and \ref{DFT}j) is consistent with a linear Stark effect, which scales proportionally with the field. We note that the charge rearrangement mechanism is valid only below $T_N$ where the Mn spins are antiferromagnetically ordered, allowing for coherent charge transfer. On the contrary, above $T_N$, due to lack of spin order, contributions from randomly oriented Mn spins will average to zero. This is echoed in our experimental observation that no coherent oscillations were detected above $T_N$ (90 K, Supplementary Fig. 6).

\subsection*{Excitation of phonon modes}
First-principles phonon calculations reveal a low-frequency $B_g$ mode at 1.7 THz that has been theoretically predicted but not reported in prior experimental studies. We identify the 1.7 THz peak in our experimental FFT spectrum as this mode, which involves out-of-plane displacements of Mn atoms with opposite spin orientations moving in opposite directions. Its excitation arises from changes in Mn-–Mn bond hybridization that reduce the magnetic moment and enhance on-site electronic repulsion, thereby generating out-of-plane forces that drive the Mn atoms apart. The excitation of the 1.7 THz mode is further corroborated by our constrained DFT calculations. Specifically, we employed the ground-state wavefunctions (obtained in the absence of the THz field) as the reference basis, and then characterized how the strong THz field modulates these wavefunctions within the constrained DFT framework. We found that the modulation of valence state could be described approximately as mixing of 0.2\% of conduction band state, primarily minority-spin Mn $d$ orbitals, to almost completely occupied (99.8\%) valence band. This setup captures essential features of the electronic response induced by a THz pulse -- namely, hybridization between valence and conduction states, and a reduction in the magnitude of the local magnetic moments.

On the other hand, the polarization-dependent field-induced change in exchange interaction $J$ preferentially couples to the  4.5 THz phonon modes via magnetoelastic interaction. This is supported by previous studies that report changes in these modes across the N\'eel temperature, suggestive of their sensitivity to the magnetic exchange parameter $J$ \cite{vaclavkova2020,sun2019}.

\subsection*{Symmetry of the 1.7 THz and 4.5 THz modes}
As displayed in Fig. \ref{fig:2}d, the 4.5 THz mode exhibits distinct symmetry depending on the THz polarization. When the THz field is aligned along the lobes of the SHG-RA pattern ($E_\perp$), a six-fold symmetry is observed (dark blue and orange curves). On the other hand, a two-fold pattern appears (purple and red curves) when the field is polarized along the nodes ($E_\parallel$). We associate this difference in the FFT spectra of the pump–probe SHG signal at 4.5 THz with anisotropic modifications in the hybridization of the six Mn–-S bonds surrounding each Mn atom. To gain further insights, we analyze the time-dependent quantity $D_{R',R}(t)=\sum_{\alpha,\beta}|\rho_{\alpha R',\beta R}(t)|^2$, which quantifies the bonding strength between the Mn site and its six neighboring S atoms computed from the off-site elements $\rho_{\alpha R',\beta R}(t)$ of the single-particle reduced density matrix. The applied THz field induces highly asymmetric changes in $D_{R_{\mathrm{Mn}}, R_{\mathrm{S}}}(t)$ among the six Mn--S bonds surrounding each Mn atom (Supplementary Fig. 11), with the response strongly dependent on the field polarization. These asymmetries in bond strength exert directional forces on the atoms, selectively exciting specific phonon modes.

We attribute the observed pump polarization dependence of the 4.5~THz mode to the selective excitation of IR- and Raman-active phonons with different symmetries. The five phonons near 4.5 THz show $A_g$, $B_g$, $A_u$ and $B_u$ symmetries. With $E_{\parallel}$ applied, the $A_g$ and $A_u$ modes dominate the response (Figure \ref{4d5}a), strongly altering $D_{R_{\mathrm{Mn}}, R_{\mathrm{S}}}(t)$ along a single Mn--S bond direction, while the bonds in other two directions remain largely unaffected. This manifests in a strongly two-fold symmetry pattern at 4.5 THz, in agreement with the experimental observation. By contrast, for the $E_\perp$ excitation, the 4.5 THz $B_g$ and $B_u$ modes are excited (Figure \ref{4d5}b). These modes induce modifications in $D_{R_{\mathrm{Mn}}, R_{\mathrm{S}}}(t)$ along all three Mn--S directions, consistent with the experimentally observed uniform six-fold response. 

At 4.5 THz, the eigenvectors of the IR-active $A_u$ and $B_u$ modes feature pronounced displacements of P atoms, whereas the Raman-active $A_g$ and $B_g$ modes display minimal phosphorus involvement (Extended Data Fig. \ref{4d5}a--b). However, according to our calculations, the SHG response in MnPS$_3$ is primarily sensitive to the Mn--S bond current. This, together with the small THz pump spectral weight around 4.5 THz (see Extended Data Fig. \ref{SHGexeriment}), suggests that the 4.5 THz response in tr-SHG measurement contains comparable contribution from both the IR-active modes directly excited by the THz field and Raman-active modes excited via field-induced charge transfer mechanism. 

The six-fold symmetry of the 1.7 THz mode observed in the experiment arises from out-of-plane displacements of Mn atoms with opposite spins moving in opposite directions. This motion alternately modulates three of the six Mn–S bonds around each Mn site. The combined effect on hybridization at the two opposite-spin Mn sites yields a symmetric pattern across all six directions, producing the six-fold symmetry observed in the FFT signal.

\subsection*{Long-lived changes and magnetic phase transition}
For stronger THz field $|E_{THz}|$, our simulations predict long-lived changes in the local charge densities and magnetic moments, which persist even after the electric field is turned off. These changes in the local charge densities and magnetic moments scale nonlinearly with $|E_{THz}|$ as presented in Fig. \ref{fig:4}g. Extended Data Fig. \ref{4d5}c--d plots the simulated temporal evolution of net magnetic moment as a function of field for different pump polarizations. Especially, in the $E_\parallel$ case, a finite net magnetization remains after the pulse (Extended Data Fig. \ref{4d5}c), hinting at a transition from the antiferromagnetic (AFM) to a ferrimagnetic (FiM) phase, characterized by unequal magnetic moments at opposite-spin Mn sites. These long-lived changes and the field-induced phase transition suggest that the system evolves through nonlinear and nonperturbative dynamics, which explains the experimentally observed rectification (offset) in the pump-probe signal. The excited electronic subsystem partially relaxes into specific phonon modes, as discussed in the “Excitation of phonon modes” section, by transferring energy via electron–phonon coupling.

We distinguish between the instantaneous charge transfer from S to Mn, which scales linearly with the THz field (Extended Data Fig. \ref{DFT}j) and reflects transient hybridization-driven redistribution during the pulse, and the long-lived offset observed in the SHG response, which scales nonlinearly with field. The offset represents the amount of charge that remains after the field is gone and builds up over the course of the excitation, indicating a rectified, cumulative effect rather than a direct linear response.

\subsection*{SHG predicted by the tight-binding model}
The AFM bulk MnPS$_3$ lacks inversion symmetry, and exhibits SHG and other even-order harmonics. As such, we employ the rt-TDDFT method based on the above tight-binding model to simulate the SHG.  The electromagnetic field effect is incorporated utilizing the Peierls substitution method \cite{Peierls1933} where the hopping amplitude is multiplied by a phase factor. We define a instantaneous current density $\mathcal{J}(t)$ on each bond connecting atoms $R$ and $R'$ where $R,R'\in \{R_{Mn},R_{S}\}$: 
\begin{eqnarray}
\mathcal{J}(t) &=&  \sum_n f_n \sum_{\langle R, R' \rangle} \sum_{\sigma} \sum_{\alpha, \beta} \frac{i}{\hbar} \Big( 
\psi^*_{\sigma, \alpha, R, n} (t) \, t^{hop}_{\alpha, \beta, R, R'} \, \psi_{\sigma, \beta, R', n} (t) \nonumber \\
&& \hspace{4em} - \psi_{\sigma, \alpha, R, n} (t)\, t^{hop}_{\alpha, \beta, R, R'} \, \psi^*_{\sigma, \beta, R', n} (t) 
\Big) d_{R-R'}.
\label{current_operator}
\end{eqnarray}

Here, $d_{R-R'}=\frac{\vec{R}-\vec{R}'}{|\vec{R}-\vec{R}'|}$ is the unit vector along the direction joining sites $R$ and $R'$. The total current in a particular spatial direction is obtained by summing the contributions from all bonds, each weighted by the component of the displacement vectors between the sites in that direction. For a probe optical field with frequency $\omega$, SHG yields are computed by analyzing the Fourier components of the current density $\mathcal{J}(t)$, with longitudinal ($\parallel E_{\omega}$) and transverse ($\perp E_{\omega}$) components taken along the directions parallel and perpendicular to the field, respectively. The SHG yield can be written as: $I_{SH}=|\tilde{\mathcal{J}}(2\omega)|^2$, where $\tilde{\mathcal{J}}$ is the Fourier transform of current density $\mathcal{J}(t)$. Two different field strengths were tested for the simulations: 0.005~V/\AA{} and 0.5~V/\AA{}.

To examine the polarization dependence of the SHG response, we calculated the longitudinal and transverse SHG yields for electric fields polarized along various directions relative to the Mn--Mn bond axis. Extended Data Fig.~\ref{DFT}b presents the simulated longitudinal SHG-RA pattern for bulk MnPS$_3$ in its AFM ground state. The polarization dependence of the simulated SHG-RA is consistent with previous experimental observations. The SHG intensity shows maxima every $60^\circ$, reflecting the approximate three-fold rotational symmetry of the $D_{3h}$ point group. In contrast, transverse SHG yields were weaker by several orders of magnitudes, congruent with the experimental results.

We now investigate how phonon modes influence the SHG-RA pattern by introducing atomic displacements corresponding to select low-frequency phonons into the equilibrium DFT structure. For each displaced configuration, we extract updated hopping parameters for the tight-binding model, optimize the electronic structure of bulk MnPS$_3$, and compute the SHG-RA pattern using rt-TDDFT. For Raman-active modes, the unequal intensities of mirror-related lobes indicate mirror symmetry breaking as depicted in Extended Data Fig.~\ref{DFT}c. Contrarily, for the IR-active phonon modes, the SHG-RA lobes become asymmetric, distorting the six-fold symmetry (Extended Data Fig.~\ref{DFT}d). 

To explore how phonon mode amplitude impacts the SHG yield, we calculate the SHG signal along one of the lobe directions in the ground-state SHG-RA pattern across a range of mode amplitudes. Figure~\ref{fig:4}b in the main text displays the corresponding relative change in SHG signal.

\section*{Data availability}
The data presented in this manuscript can be available from the corresponding authors upon reasonable request. Correspondence should be addressed to Sheikh Rubaiat Ul Haque, Tony F. Heinz or Aaron M. Lindenberg. 

%


\section*{Acknowledgements}
We thank Keith A. Nelson, Richard D. Averitt, Eugene Demler, Angel Rubio, Peter E. Blöchl, David A. Reis, Dragan Mihailovic, Liuyan Zhao, Alfred Zong, Shengxi Huang, Oleg Shpyrko, Roopali Kukreja, Keshav M. Dani, Ralph H. Page, Yuan Ping, Yuki Kobayashi, Jiaojian Shi, Qitong Li and Christian Heide for fruitful discussions. The experimental effort was  primarily supported by the US Department of Energy (DOE), Office of Science, Office of Basic Energy Sciences (BES), Materials Sciences and Engineering Division under contract DE-AC02-76SF005. Use of the facilities at Linac Coherent Light Source (LCLS), SLAC National Accelerator Laboratory, is supported by the US Department of Energy, Office of Science, Office of Basic Energy Sciences under Contract No. DE-AC02-76SF00515. Theory and simulation were supported by the Computational Materials Sciences Program funded by the US Department of Energy, Office of Science, Basic Energy Sciences, Materials Sciences and Engineering Division. Additional data analysis and interpretation was provided by the User Program of the Molecular Foundry, supported by the Office of Science, Office of Basic Energy Sciences, of the US Department of Energy under Contract No. DE-AC02-05CH11231. This research used resources of the National Energy Research Scientific Computing Center (NERSC), a Department of Energy User Facility using NERSC award BES-ERCAP0032784. M.J.C, M.C.H, and J.D.K. were supported by the Department of Energy, Laboratory Directed Research and Development program at SLAC National Accelerator Laboratory, under contract DE-AC02-76SF00515. L.B. and V.K. are supported by DOE-BES through the award DE-SC0002613. The National High Magnetic Field Laboratory (NHMFL) acknowledges support from the US-NSF Cooperative Agreement grant no. DMR-2128556 and the State of Florida. D.S. also acknowledges the support from the US Department of Energy (No. DE-FG02-07ER46451) for Raman spectroscopy measurements.

\section*{Author contribution}
S.R.U.H. conceived the project along with A.M.L. and T.F.H. V.K. performed the sample growth under the supervision of L.B. A.C.Z and M.T. characterized the samples for Raman measurements. I.J.S. measured the Raman spectra under the guidance of D.S. The infrared absorption spectrum was measured by M.O. S.R.U.H., H.W. and S.S.P. carried out the static SHG characterization of the sample. M.C.H. and M.J.C. designed and built the THz pump experiments. S.R.U.H. and M.J.C. performed the time-resolved measurements and analyzed the data with contribution from C.S. J.D.K. provided further support in data analysis. S.R., J.B.H. and C.J.C. performed the first-principles density functional theory (DFT) calculations under the guidance of T.O. and F.H.d.J., with support from L.Z.T. A.M.L. and T.F.H. supervised the project. S.R.U.H., S.R., J.B.H. and A.M.L. wrote the manuscript with input from all the authors. 

\section*{Competing interests}
The authors declare no competing interests.

\newpage
\setcounter{figure}{0}
\renewcommand{\figurename}{\textbf{Extended Data Fig.}}
\renewcommand{\thefigure}{\arabic{figure}}
\begin{figure} [hbt!]
    \centering
    \includegraphics[width=\linewidth]{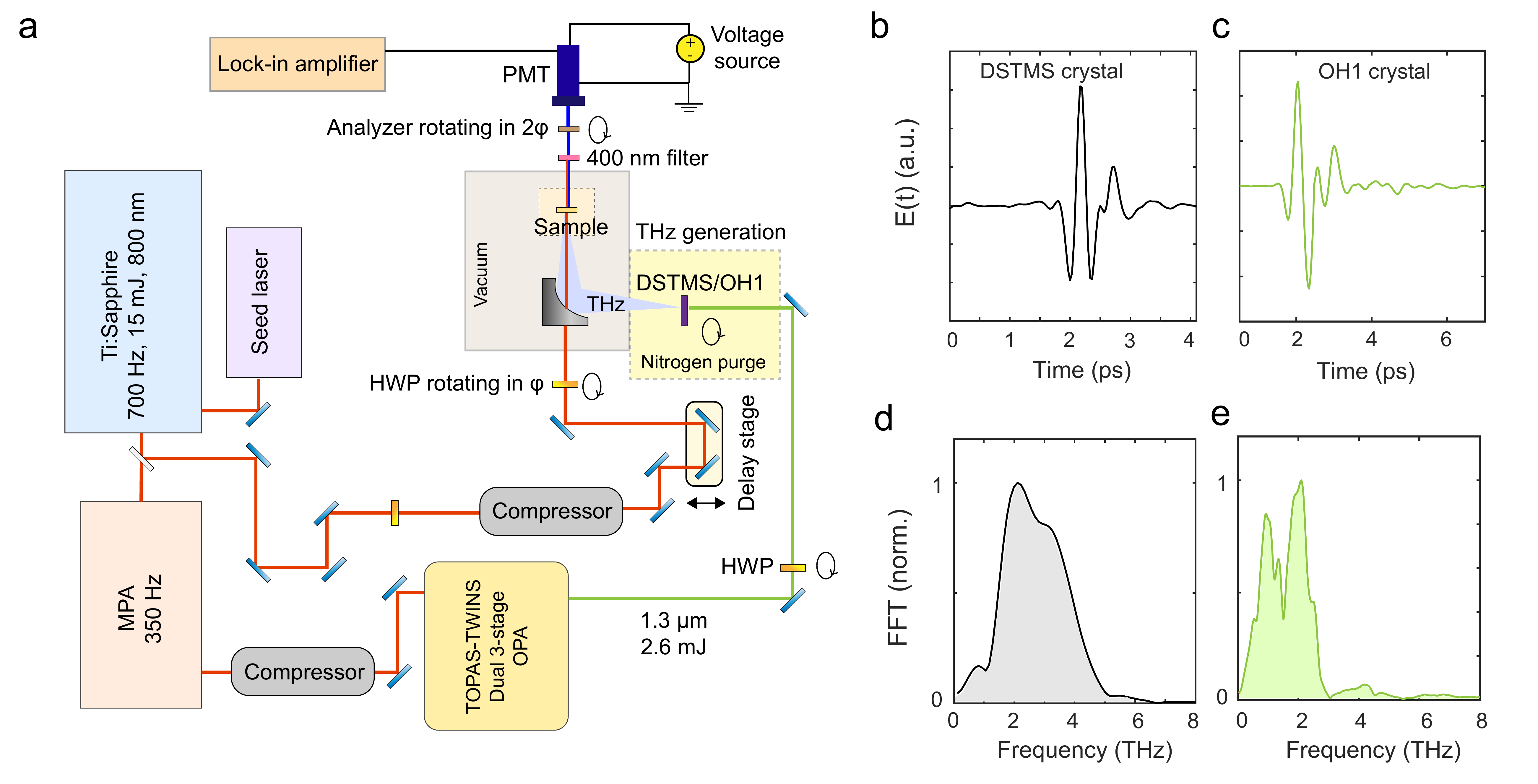}
    \caption{\textbf{Schematic of the experimental configuration. a}, Experimental setup for  strong-field THz pump - SHG probe spectroscopy. MPA: multi-pass amplifier, OPA: optical parametric amplifier, HWP: half-wave plate, PMT: photomultiplier tube. \textbf{b--e}, Electro-optic sampling (EOS) profiles and the corresponding FFT spectra of THz pump pulses generated from DSTMS (\textbf{b, d}) and OH1 (\textbf{c, e}) crystals. }
    \label{SHGexeriment}
\end{figure}

\newpage
\begin{figure} [hbt!]
    \centering
    \includegraphics[width=\linewidth]{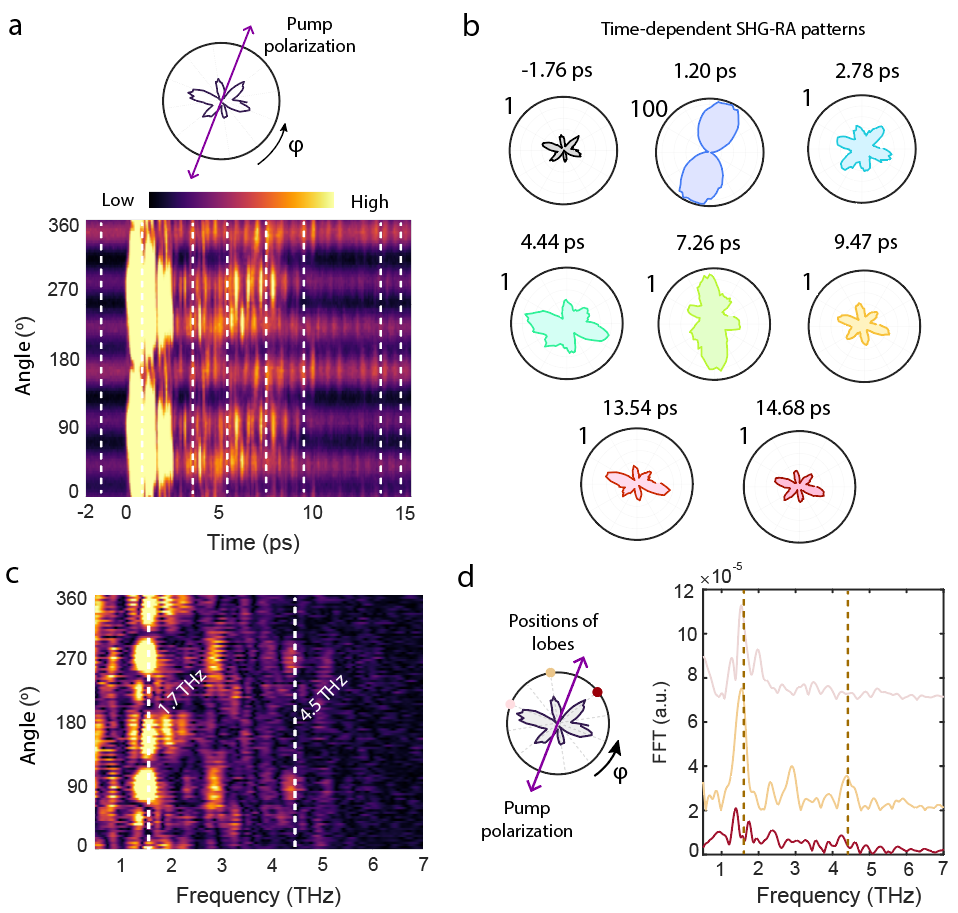}
    \caption{\textbf{Time- and angle-resolved SHG dynamics upon strong-field THz excitation (polarization 2). a}, Pump-induced SHG as a function of pump-probe delay time $t$ and polarizer analyzer angle $\varphi$ at 9 K. THz pump with field strength of $\sim$ 500 kV/cm is polarized along a node of the static SHG-RA pattern (purple arrow). The later time dynamics is dominated by a long-lived oscillatory signal, suggesting coherent phonons. \textbf{b}, SHG-RA patterns at different pump-probe delays, obtained from vertical linecuts in \textbf{a} (dashed white lines). The radial scale of each polar plot is normalized, with the outer circle corresponding to an amplitude of 1. \textbf{c}, Fourier transform spectrum of the pump-probe signal, revealing the coherent modes at 1.7 THz and 4.5 THz. The 4.5 THz mode shows a $180^\circ$ periodicity \textbf{d}, Horizontal linecuts showing FFT spectra at lobe positions (indicated by colored circles in the SHG-RA pattern). The linecuts are vertically offset for better visibility.}
    \label{SHG_pol2}
\end{figure}

\newpage
\begin{figure} [hbt!]
    \centering
    \includegraphics[width=\linewidth]{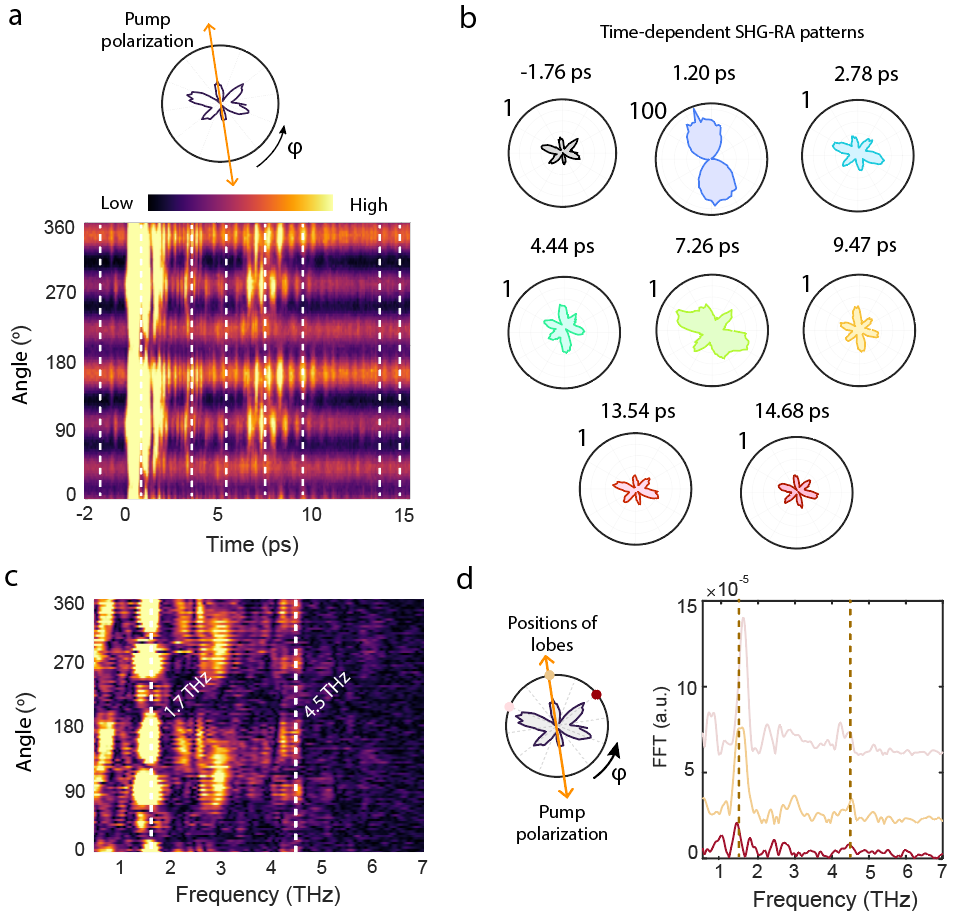}
    \caption{\textbf{Time- and angle-resolved SHG dynamics upon strong-field THz excitation (polarization 3). a}, Pump-induced SHG as a function of pump-probe delay time $t$ and polarizer analyzer angle $\varphi$ at 9 K. THz pump with field strength of $\sim$ 500 kV/cm is polarized along a lobe of the static SHG-RA pattern (orange arrow). The later time dynamics is dominated by a long-lived oscillatory signal, suggesting coherent phonons. \textbf{b}, SHG-RA patterns at different pump-probe delays, obtained from vertical linecuts in \textbf{a} (dashed white lines). The radial scale of each polar plot is normalized, with the outer circle corresponding to an amplitude of 1. \textbf{c}, Fourier transform spectrum of the pump-probe signal, revealing the coherent modes at 1.7 THz and 4.5 THz. \textbf{d}, Horizontal linecuts showing FFT spectra at lobe positions (indicated by colored circles in the SHG-RA pattern). The linecuts are vertically offset for better visibility. }
    \label{SHG_pol3}
\end{figure}

\newpage
\begin{figure} [hbt!]
    \centering
    \includegraphics[width=\linewidth]{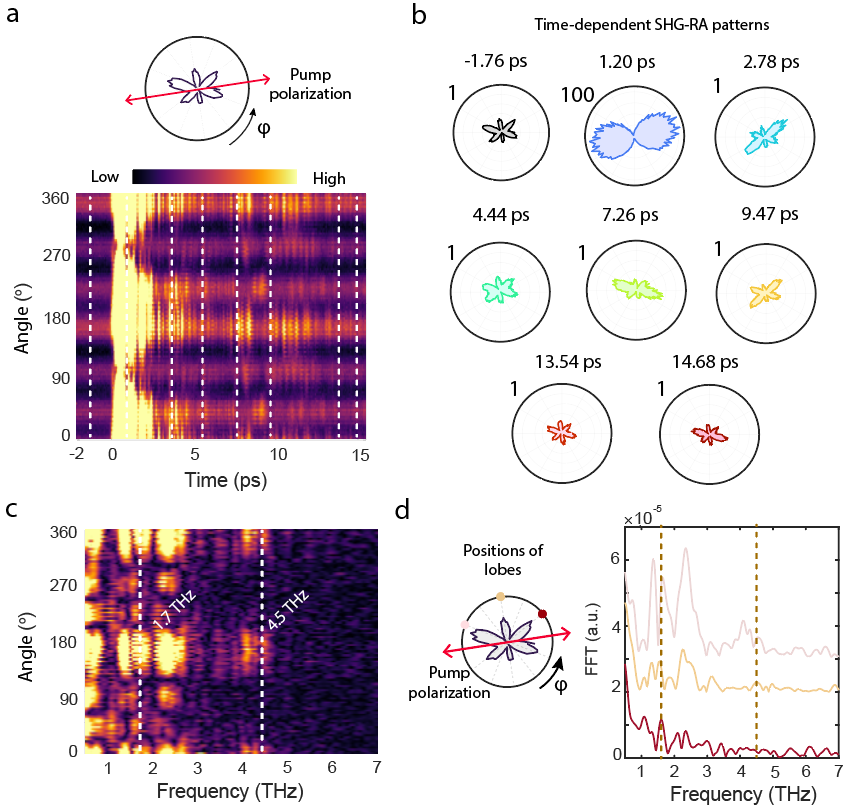}
    \caption{\textbf{Time- and angle-resolved SHG dynamics upon strong-field THz excitation (polarization 4). a}, Pump-induced SHG as a function of pump-probe delay time $t$ and polarizer analyzer angle $\varphi$ at 9 K. THz pump with field strength of $\sim$ 500 kV/cm is polarized along a node of the static SHG-RA pattern (red arrow). The later time dynamics is dominated by a long-lived oscillatory signal, suggesting coherent phonons. \textbf{b}, SHG-RA patterns at different pump-probe delays, obtained from vertical linecuts in \textbf{a} (dashed white lines). The radial scale of each polar plot is normalized, with the outer circle corresponding to an amplitude of 1. \textbf{c}, Fourier transform spectrum of the pump-probe signal, multiple prominent peaks are observed. The 4.5 THz mode shows a $180^\circ$ periodicity. \textbf{d}, Horizontal linecuts showing FFT spectra at lobe positions (indicated by colored circles in the SHG-RA pattern), showing coherent modes predominantly at 1.7 THz and 4.5 THz. The linecuts are vertically offset for better visibility.}
    \label{SHG_pol4}
\end{figure}

\newpage
\begin{figure} [hbt!]
    \centering
    \includegraphics[width=\linewidth]{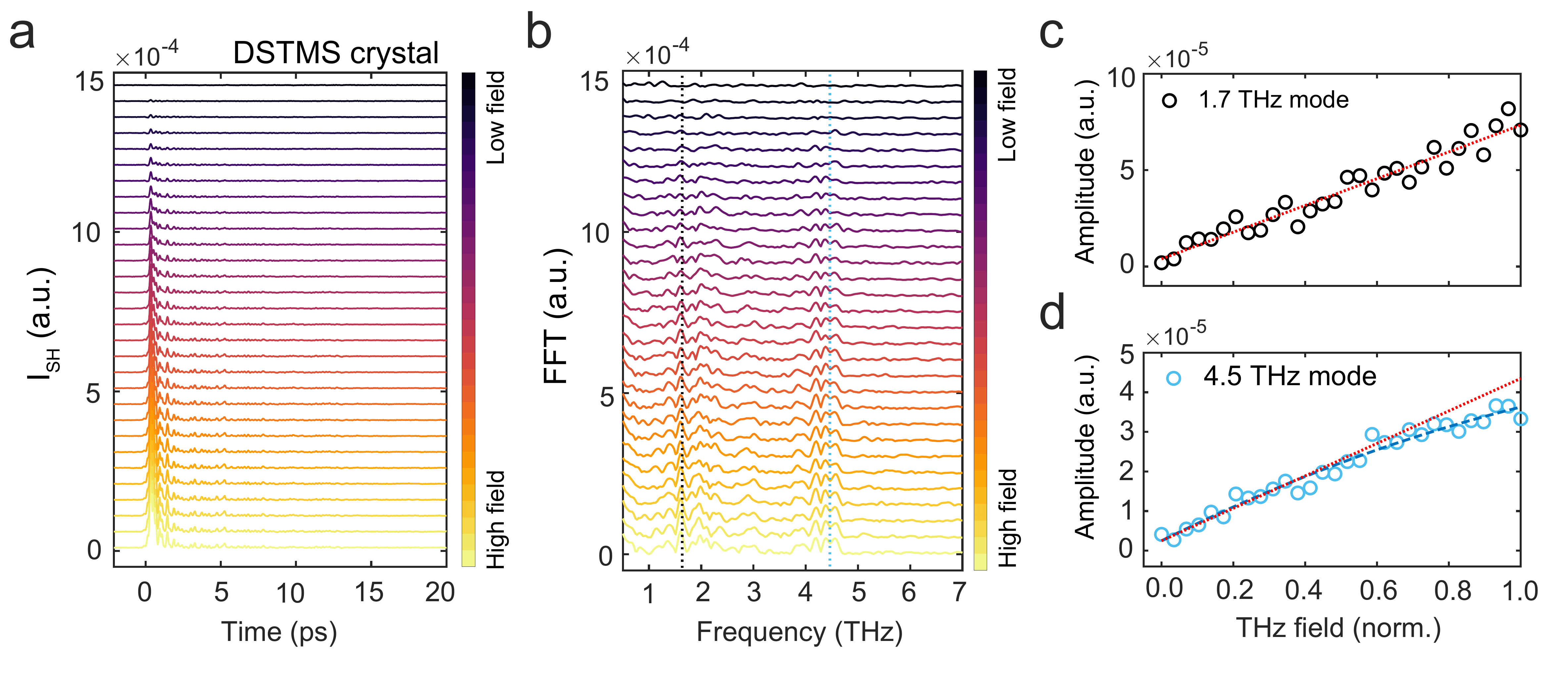}
    \caption{\textbf{tr-SHG dynamics as a function of THz pump field. a}, The tr-SHG dynamics along a lobe as a function of THz field generated from a DSTMS crsytal (peak field $\sim 500$ kV/cm) at 9 K. \textbf{b}, Corresponding FFT spectra showing pronounced peak features at 1.7 THz and 4.5 THz (vertical dashed lines). The signals are vertically offset for clarity. \textbf{c--d}, Mode amplitude plotted as a function of normalized pump field for 1.7 THz (\textbf{c}, black open circles) and 4.5 THz (\textbf{d}, blue open circles) modes. Both modes exhibit predominantly linear scaling with field. Red and blue dashed lines indicate linear and sublinear fits, respectively.}
    \label{field_dependence}
\end{figure}

\newpage
\begin{figure} [hbt!]
    \centering
    \includegraphics[width=\linewidth]{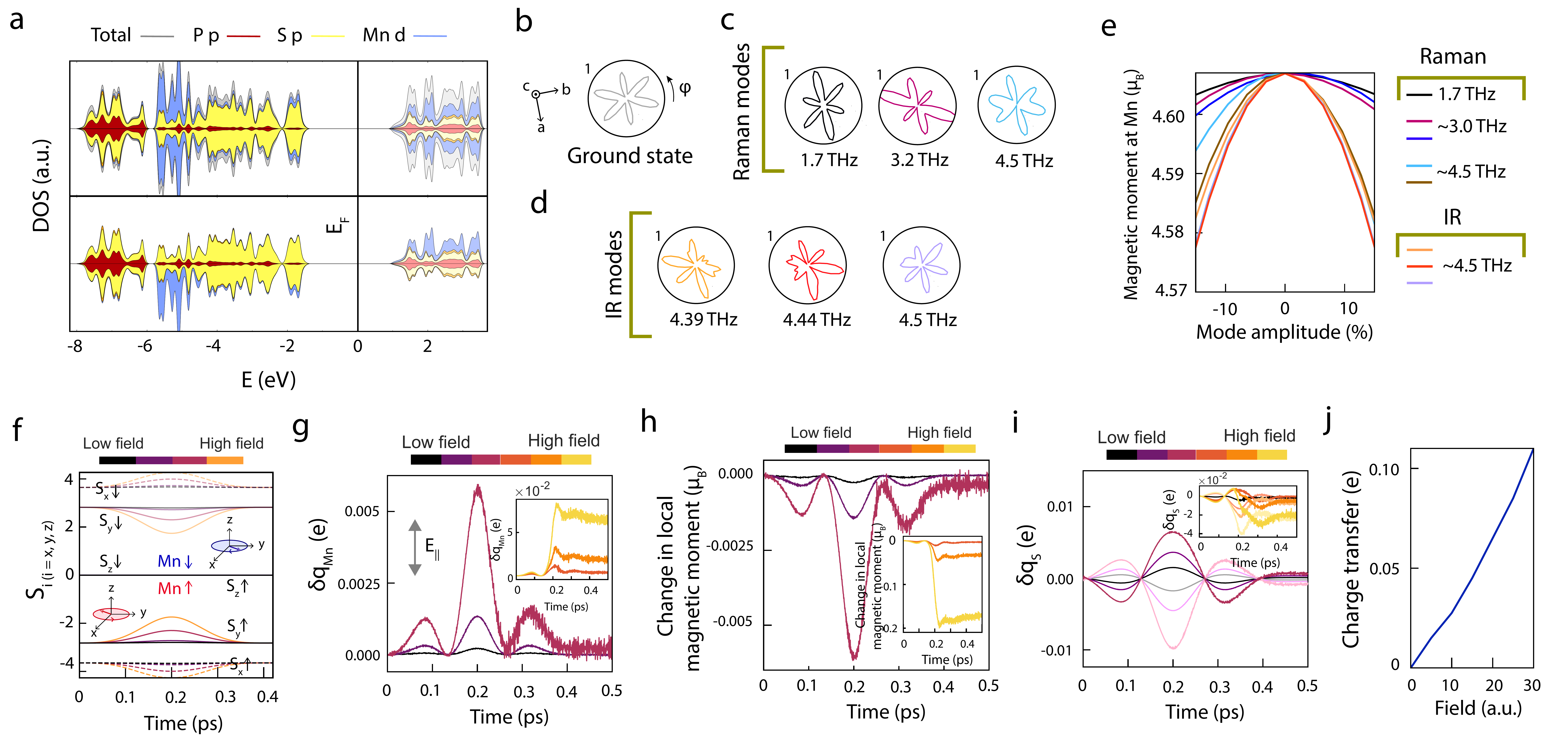}
    \caption{\textbf{DFT calculations.} \textbf{a}, Density of states (DOS) of bulk MnPS$_3$. Top: total DOS (gray) projected onto Mn $d$ (blue), S $p$ (yellow), and P $p$ (red) orbitals. Bottom: same as the top panel, but with projections restricted to Mn ions with up spin. The vertical solid line denotes the Fermi level. \textbf{b}, Ground state SHG-RA pattern reconstructed via tr-TDDFT using the $p - d$ model. \textbf{c--d}, Computed SHG-RA patterns combining rt-TDDFT and frozen phonon methods for coupling with Raman- (\textbf{c}) and IR-active (\textbf{d}) modes. \textbf{e}, Variation of average magnetic moment at Mn sites with phonon mode amplitude, calculated from the tight-binding model. The phonon amplitudes are given as \% of the equilibrium Mn–S bond length. IR- and Raman-active phonons at $\sim$4.5 THz exhibit notable modulation of magnetic moment. \textbf{f}, Evolution of individual spin components $S_x$, $S_y$, and $S_z$ for Mn sites with opposite spins during the THz magnetic field $B_{THz}$. The direction of $B_{THz}$ is along the $z$-axis. For clarity, solid and transparent lines are used for opposite spin orientations. The insets illustrate the precession of opposite Mn spins (red and blue arrows) about the $z$-axis. \textbf{g}, Dynamical simulations of the charge density at Mn atoms under a single-cycle 2.5 THz electric field. The inset displays the response at higher fields, where a residual offset remains after the THz pulse. \textbf{h}, dynamics of average local magnetic moments at Mn sites for lower (main) and higher (inset) field strengths, revealing decrease in average magnetic moment. \textbf{i}, Modulation of charge density at S sites upon excitation. Solid and transparent colors represent S atoms on top and bottom of Mn layers. \textbf{j}, Charge transfer from S to Mn as a function of field in the low-field limit, showing linear scaling.}
    \label{DFT}
\end{figure}

\newpage
\begin{figure} [hbt!]
    \centering
    \includegraphics[width=\linewidth]{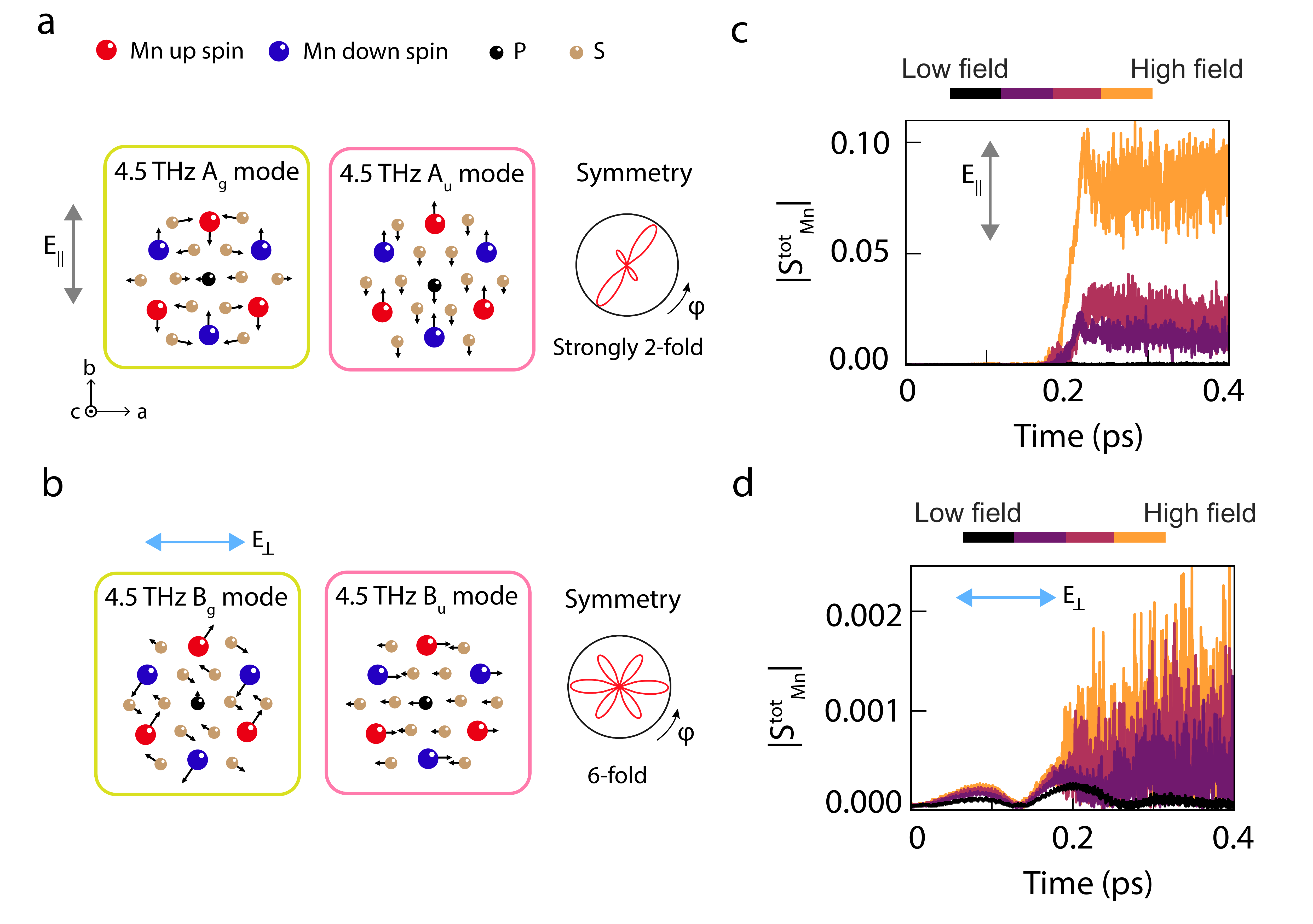}
    \caption{\textbf{THz polarization dependence.} \textbf{a}, THz field-induced Atomic displacements corresponding to the Raman-active $A_g$ and IR-active $A_u$ modes for $E_\parallel$ (gray arrow) case. The black arrows represent the displacements of corresponding atoms. Selective excitation of these two modes leads to an experimentally observed two-lobe symmetry pattern at 4.5 THz. \textbf{b}, $E_\perp$ (blue arrow) preferentially excites the 4.5 THz Raman-active $B_g$ and IR-active $B_u$ modes, resulting in a six-fold symmetry pattern. \textbf{c--d}, rt-TDDFT simulations showing the evolution of net magnetic moment under a varying THz field for $E_\parallel$ (\textbf{c}) and $E_\perp$ (\textbf{d}). Compared to the $E_\perp$ polarization, a net magnetic moment persists for  $E_\parallel$ pump, indicating an AFM-FiM transition.}
    \label{4d5}
\end{figure}

\newpage
\begin{table}[!ht]
\renewcommand{\thetable}{\arabic{table}}
\makeatletter
\renewcommand{\fnum@table}{\textbf{ Extended Data Table~\thetable}}
\makeatother
\begin{center}
\begin{tabular}{|lrl|lrl|}
\hline
\hline
  ($\epsilon^o_{p\uparrow}+\epsilon^o_{p\downarrow}$)/2       & 6.15 & eV
&   ($\epsilon^o_{d\uparrow}+\epsilon^o_{d\downarrow}$)/2            & 7.66 & eV \\
 $\epsilon^o_{d\uparrow}-\epsilon^o_{d\downarrow}$        & 7.016 & eV
& $\Delta E_{FM-AFM}$        & 0.10 & eV \\
$\epsilon_p$           & 0 & eV  
& $\epsilon_d$ & -1.51 & eV \\
  $U$           & 3.01 & eV  
& $J_{xc}$ & 1.00  & eV \\ 
  $A_{pd\sigma}$  & 0.90 & eV 
&  $A_{pd\pi}$ & -0.45 & eV \\
\hline
\end{tabular}
\caption{\label{tab:t1} Parameters for the tight-binding model. For detailed information
on the extraction of these model parameters from the
first-principle studies \cite{Sotoudeh2017}.}
\end{center}
\end{table}

\newpage
\begin{table}[!ht]
\renewcommand{\thetable}{\arabic{table}}
\makeatletter
\renewcommand{\fnum@table}{\textbf{Extended Data Table~\thetable}}
\makeatother
\begin{center}
\begin{tabular}{|c|c|c|c|}
\hline
\hline
Mode & Frequency (THz) & Symmetry & IR activity \\
\hline
4  & 2.0519 & $B \ (B_g)$ & 0.0000 \\
5  & 3.0134 & $B \ (B_g)$ & 0.0000 \\
6  & 3.0361 & $A \ (A_g)$ & 0.0000 \\
7  & 4.1364 & $A \ (A_g)$ & 0.0001 \\
8  & 4.1708 & $B \ (B_g)$ & 0.0001 \\
9  & 4.3633 & $B \ (B_u)$ & 0.4813 \ (z-polarized) \\
10 & 4.4734 & $A \ (A_u)$ & 1.5398 \ (x-polarized $E_{\parallel}$) \\
11 & 4.4888 & $B \ (B_u)$ & 1.7042 \ (y-polarized $E_{\perp}$) \\
\hline
\end{tabular}
\caption{\label{tab:phonon-calc} Vibrational mode frequencies (in THz), symmetries, and IR activities for modes 4–11. IR activity $= \sum_i e^2|Z^{(\nu)}_i|^2$, where $Z^{(\nu)}$ is the Born effective charge in the phonon mode basis. Units are in (Debye/\text{\AA})$^2$/a.m.u.}
\end{center}
\end{table}

\newpage
\section*{Supplementary Materials}
\subsection*{Infrared spectrum} 
Far-infrared measurements were performed at the National High Magnetic Field Laboratory, using a Bruker Vertex 80v vacuum Fourier Transform Infrared (FTIR) spectrometer coupled to a vertical-bore superconducting 17.5 T magnet. Transmission spectra of a single crystal of MnPS$_3$ were recorded in the spectral range from 10 cm$^{-1}$ to 205~cm$^{-1}$, with a resolution of 0.3~cm$^{-1}$.

Broadband far-infrared radiation, generated by a mercury lamp, was guided through an evacuated beamline connecting the spectrometer to the top of a brass lightpipe. A 90$^\circ$ off-axis parabolic mirror located at the bottom of the lightpipe directed the beam toward a second, identical confocal mirror, which collimated the radiation through a short secondary lightpipe ending at a Si bolometer detector. The sample, mounted at the focal point of the mirrors, fully covered a clear aperture of 3~mm and was maintained at a fixed temperature of approximately 5.5~K.

\setcounter{figure}{0}
\renewcommand{\figurename}{\textbf{ Supplementary Fig.}}
\renewcommand{\thefigure}{\arabic{figure}}
\begin{figure} 
    \centering
    \includegraphics[width=\linewidth]{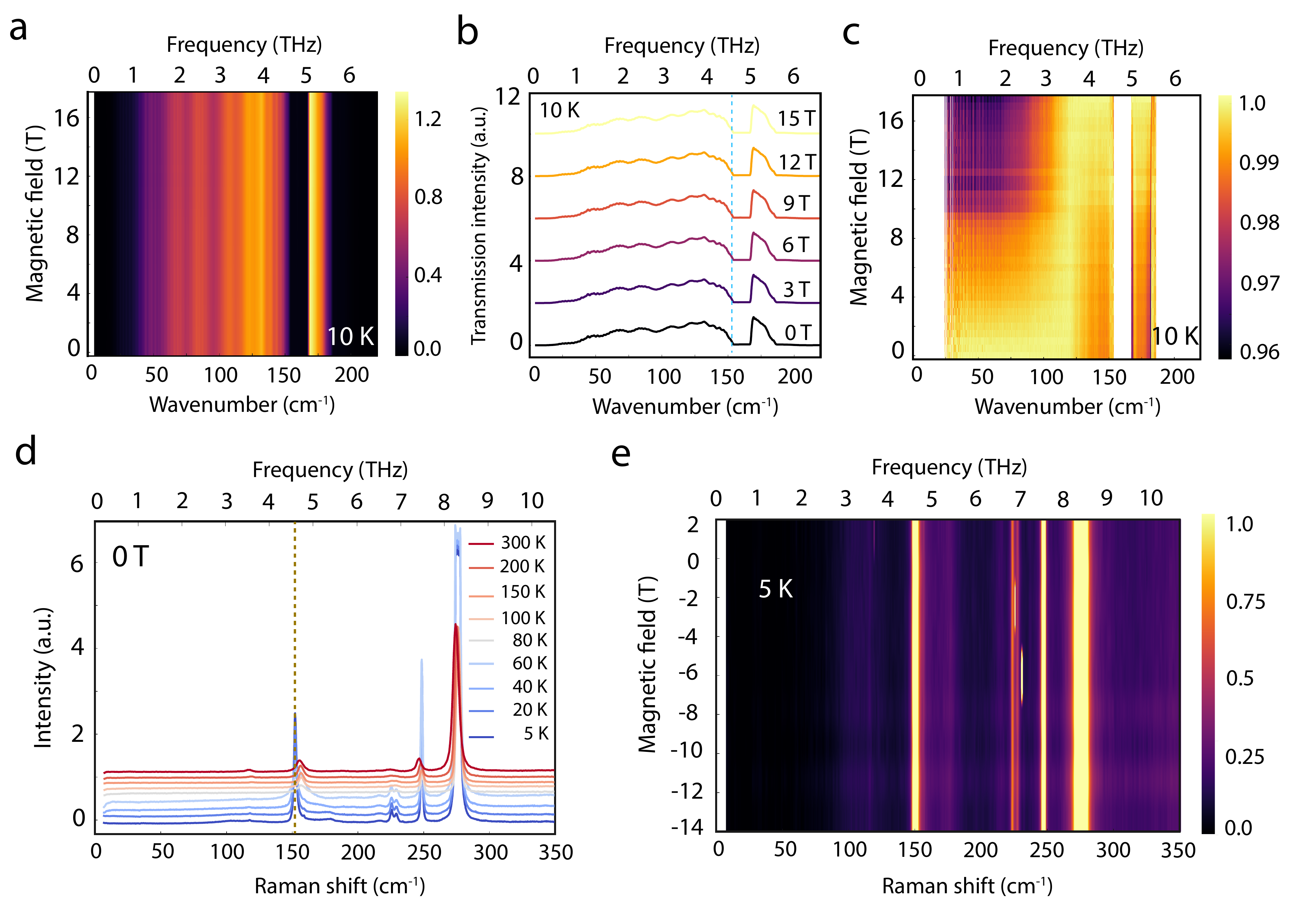}
    \caption{\textbf{Infrared and Raman spectra. a}, 2D colormap of IR spectra at 10 K as a function of applied in-plane magnetic field, showing a sharp dip at $\sim 4.5$ THz (150 cm$^{-1}$). \textbf{b}, Horizontal linecuts at select magnetic fields, the blue dashed line at 4.5 THz represents a phonon. Vertical offsets are applied for better visibility. \textbf{c}, Magnetic field-dependent spectra (normalized with the spectrum at 0 T) showing negligible field-induced changes. \textbf{d}, Temperature-dependent Raman spectra. The 4.5 THz phonon is marked in brown dashed line. \textbf{e}, Raman spectra at 5 K as a function of in-plane magnetic field shown in a 2D colormap. No spectral change due to magnetic field was recorded. }
    \label{raman_IR}
\end{figure}

The magnetic field was applied in the plane of the sample and perpendicular to the propagation direction of the incident radiation (Voigt geometry). A cold low-pass filter blocked infrared radiation above 200~cm$^{-1}$. The transmission spectra were acquired in 0.5~T magnetic field increments and are presented as an intensity colormap in Supplementary Fig.~\ref{raman_IR}a while corresponding horizontal line cuts, taken at every 3~T, are shown in Supplementary Fig.~\ref{raman_IR}b. A dip in transmission intensity was observed at $\sim4.5$ THz (150 cm$^{-1}$), matching the frequency of an IR-active mode. To improve the signal-to-noise ratio, each measurement was repeated twice at every field step. The reproducibility between repeated scans was better than 0.5\%, though slightly degraded near phonon peak regions due to reduced intensity.

To isolate field-dependent spectral features from field-independent background effects, each spectrum was divided by a reference spectrum, defined as the average of all measured spectra with statistical outliers removed. This normalization effectively suppressed instrumental artifacts and yielded a flat baseline, while preserving field-responsive changes.

The resulting normalized spectra were presented as a colormap in Supplementary Fig. \ref{raman_IR}c, with black and yellow representing the minimum and maximum deviations from the reference, respectively. White regions indicate frequencies where the spectral reproducibility exceeded 1\%, typically coinciding with diminished transmission intensity. A broad purple-black band below 100~cm$^{-1}$ reflects a known $\sim$3\% reduction in source intensity under high magnetic fields in this experimental configuration.

\subsection*{Raman spectrum} 
For low-temperature Raman spectroscopy measurements at National High Magnetic Field Laboratory, a 0.6 W 532 nm laser excitation was employed with a 2400 grooves/mm grating in a 0.75 FL spectrometer. To ensure better spectral resolution over a spectral window centered at 180 cm$^{-1}$, a 30 $\mu m$ entrance slit was used. For each set of measurement, high signal-to-noise ratio was obtained upon averaging over 3 acquisitions (360 s acquisition time). Moreover, the averaging would also eliminate the cosmic ray effects. The experiments were performed under unpolarized light conditions using a polarization scrambler. To perform high-temperature ($>60$ K) measurement, the laser power and grating were changed to 0.2 W and 1200 grooves/mm, respectively. In addition, the spectral window was centered at 470 cm$^{-1}$.

Supplementary Fig. \ref{raman_IR}d plots the temperature-dependent Raman spectrum under zero magnetic field. A sharp peak is visible at 4.5 THz ($\sim 150$ cm$^{-1}$), consistent with previous studies \cite{vaclavkova2020, sun2019, kim2019mps}. Interestingly, as the temperature is lowered, a broadband feature around 3 THz ($\sim 100$ cm$^{-1}$) and a sharp feature at 3.5 THz ($\sim 117$ cm$^{-1}$) appear below 60 K, potentially suggesting an onset of inversion symmetry-breaking. Upon applying an in-plane magnetic field, no change in the Raman spectrum at 5 K was recorded as described in the 2D colormap in Supplementary Fig. \ref{raman_IR}e, ruling out the possibilities of magnon or magneto-phonon coupling.

\subsection*{Phenomenological fit of the SHG-RA patterns}
For a normally incident probe beam, the SHG intensity as a function of the analyzer angle $\varphi$ is given by the expression

\setcounter{equation}{0}
\renewcommand{\theequation}{S\arabic{equation}} 
\begin{equation}
    I_{SH}=|A cos^2(\varphi) sin(\varphi) + B sin^3(\varphi)|^2,
    \label{fit_1}
\end{equation}
where $A$ and $B$ are phenomenological fit parameters, and can be written as a linear combination of the susceptibility tensor elements: $A = \chi_{xyy} + \chi_{yxy} + \chi_{yyx}$, $B=\chi_{xxx}$. All other components of the SHG tensor vanish within the electric dipole approximation as a consequence of the mirror plane perpendicular to $y$-axis \cite{chu2020}. For convenience, we take $A$ to be a real number while $B$ is a complex quantity, i.e., $B =B_{real}+B_{im}$ \cite{chu2020, ni2021}. Using this formulation allows for a reasonable fit of the equilibrium SHG-RA patterns. However, this model breaks down for pumped SHG-RA patterns since $A$ and $B$ alone are unable to capture the THz field-induced large symmetry modulations as seen in Fig. 3d. To resolve this, we add extra terms in (\ref{fit_1}):
\begin{equation}
    I_{SH}=|A cos^2(\varphi) sin(\varphi) + B sin^3(\varphi)+C sin^2(\varphi) cos(\varphi) +D cos^3(\varphi)|^2.
    \label{fit_2}
\end{equation}

Here, $C=C_{real}+C_{im} = \chi_{yxx} + \chi_{xxy}+ \chi_{xyx}$ and $D=D_{real}+D_{im}=\chi_{yyy}$. Eq. (\ref{fit_2}) enables us to fit the nonequilibrium SHG-RA patterns as a function of pump-probe delay time (Fig. 3e). A full derivation of Eqs. (\ref{fit_1})-(\ref{fit_2}) is provided in the next section. To avoid complications, we restrict our analysis to the nonequilibrium SHG-RA patterns at time delays well beyond the TFISH response (see the phenomenological fitting movies in the Supplementary Materials). 

For the pump polarization described in Supplementary Fig. \ref{fitting}a (THz is generated from a DSTMS source), the temporal evolution of the real and imaginary parts of the fitting parameters $A, B, C$ and $D$ are displayed in Supplementary Figs. \ref{fitting}a and \ref{fitting}c, respectively. The corresponding Fourier transform spectra are shown in Supplementary Figs. \ref{fitting}b and \ref{fitting}d, revealing pronounced phonon peaks at 1.7 and 4.5 THz. These results are consistent with other pump polarizations (Supplementary Fig. \ref{fitting}e--h) as well as different THz generation crystal (OH1, Supplementary Figs. \ref{OH1_symm}--\ref{fitting_OH1}).

Alternatively, instead of fitting the data with Eq. (\ref{fit_2}), we can Fourier decompose the nonequilibrium signal in the following form:
\begin{equation}
    I_{SH}=a_0+a_2 cos(2\varphi)+a_4 cos(4\varphi)+a_6 cos(6\varphi)+b_2 sin(2\varphi)+b_4 sin(4\varphi)+b_6sin(6\varphi),
    \label{fit_3_fourier}
\end{equation}
where $a_0, a_2, a_4, a_6, b_2, b_4, b_6$ are the Fourier decomposition coefficients. Now, equating Eqs. (\ref{fit_2}) and (\ref{fit_3_fourier}) yields the following relations:
\begin{align}
a_0 &= \frac{1}{32} \Big( 2A^2 + 10|B|^2 + 2|C|^2  + 10|D|^2 + 4A\,\mathrm{Re}(B) + 4\,\mathrm{Re}(CD^*)\Big) \label{eq56}\\
a_2 &= \frac{1}{32} \Big( A^2 - 15|B|^2 - |C|^2  + 15|D|^2 - 2A\,\mathrm{Re}(B) + 2\,\mathrm{Re}(CD^*) \Big)\\
a_4 &= \frac{1}{32} \Big( -2A^2 + 6|B|^2 - 2|C|^2 + 6|D|^2 - 4A\,\mathrm{Re}(B) - 4\,\mathrm{Re}(CD^*)\Big) \\
a_6 &= \frac{1}{32} \Big( -A^2 - |B|^2 + |C|^2  + |D|^2 + 2A\,\mathrm{Re}(B) - 2\,\mathrm{Re}(CD^*)\Big) \\
b_2 &= \frac{1}{32} \Big( 6A\,\mathrm{Re}(C) + 10A\,\mathrm{Re}(D) + 10\,\mathrm{Re}(BC^*) + 6\,\mathrm{Re}(BD^*) \Big) \\
b_4 &= \frac{1}{32} \Big( 8A\,\mathrm{Re}(D) - 8\,\mathrm{Re}(BC^*) \Big) \\
b_6 &= \frac{1}{32} \Big( -2A\,\mathrm{Re}(C) + 2A\,\mathrm{Re}(D) - 2\,\mathrm{Re}(BD^*) + 2\,\mathrm{Re}(BC^*) \Big)\label{eq13}
\end{align}

\begin{figure} 
    \centering
    \includegraphics[width=\linewidth]{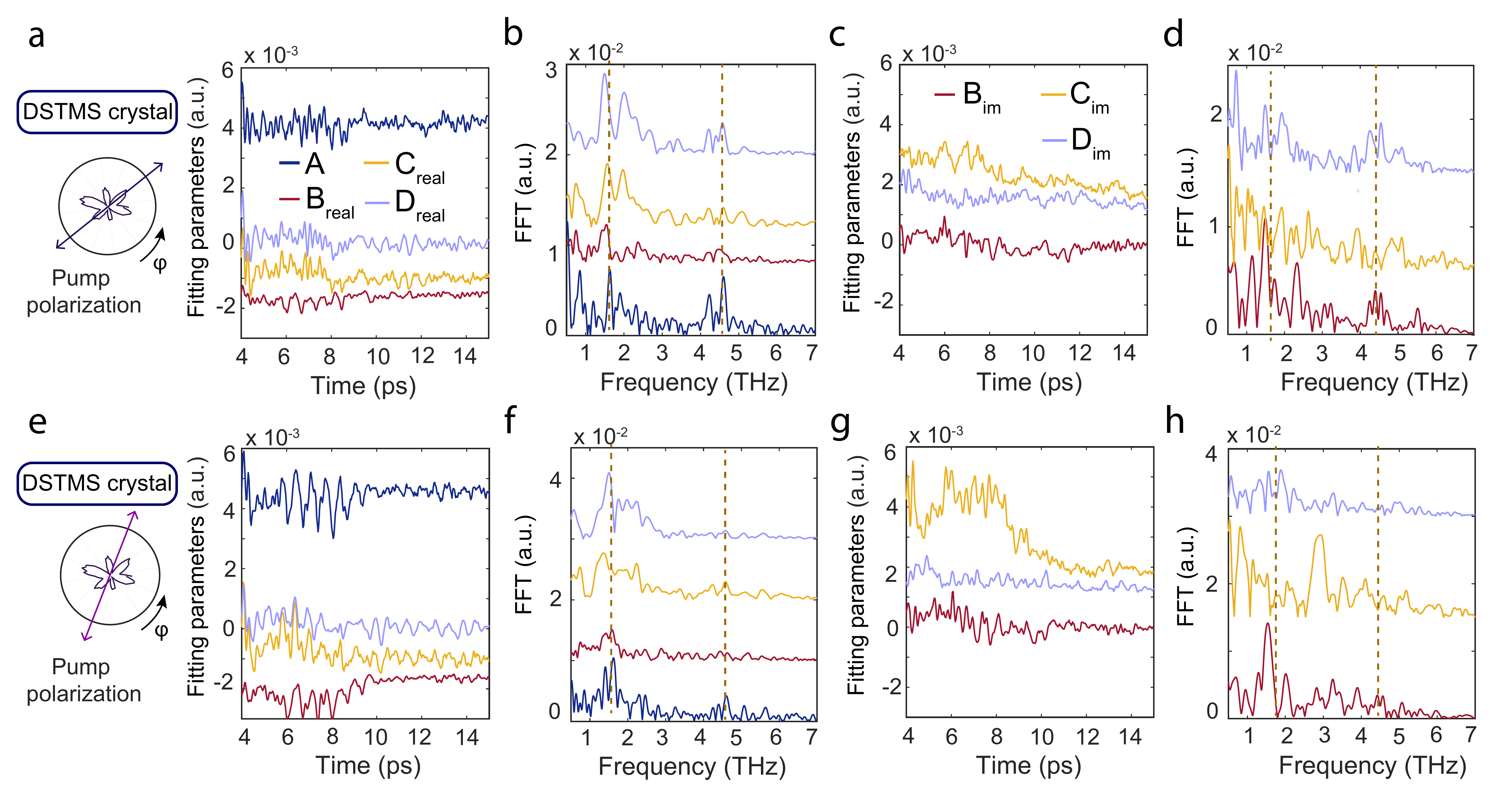}
    \caption{\textbf{Time evolution of complex fitting parameters for the SHG-RA patterns. a--d}, Real (\textbf{a}) and imaginary (\textbf{c}) parts of the SHG fit parameters $A$, $B$, $C$, and $D$, and their corresponding Fourier transforms (\textbf{b}, \textbf{d}) at 9 K. The THz pump is generated from a DSTMS crystal, and is polarized along a lobe relative to the SHG-RA pattern (dark blue arrow). Vertical dashed lines at 1.7~THz and 4.5~THz highlight dominant coherent modes. \textbf{e--h,} Same analysis with the pump polarization along a node (purple arrow), showing changes in the amplitude and frequency content of the fitted components. The FFTs are separated by vertical offsets for clarity.}
    \label{fitting}
\end{figure}

Equations (\ref{eq56})--(\ref{eq13}) imply that the Fourier coefficients $a_0, a_2, a_4, a_6, b_2, b_4, b_6$ can be reconstructed from the complex fit parameters $A, B, C, D$. For the THz pump polarized along a lobe, the Fourier coefficients, obtained directly from the data and reconstructed from the fit parameters using Eqs. (\ref{eq56})–(\ref{eq13}), are shown in Supplementary Figs. \ref{fourier_decomp}b and \ref{fourier_decomp}c, respectively. A similar analysis is presented in Supplementary Figs. \ref{fourier_decomp}e and \ref{fourier_decomp}f, where the THz pump is polarized along a node. From the figures, it is clear that the coefficients acquired from both methods exhibit strikingly similar trends, supporting the validity of our phenomenological fit analysis as well as the accuracy of the fit parameters.

\begin{figure*}
    \centering
     \includegraphics[width=140 mm]{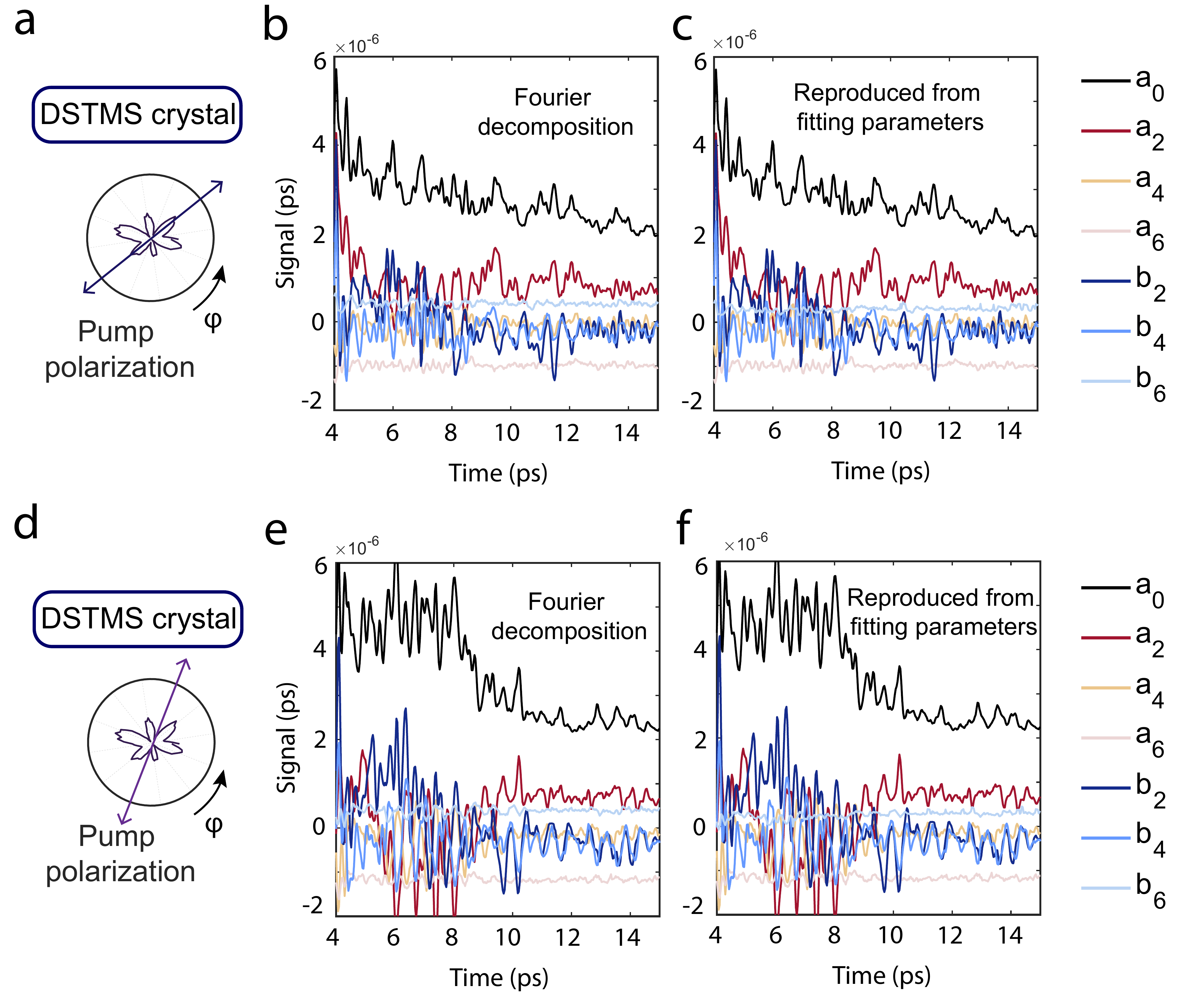}
    \caption{\textbf{Quantitative analysis of SHG-RA patterns. a}, The THz pump is polarized along a lobe of the SHG-RA pattern (dark blue arrow). The THz is generated from a DSTMS crystal. \textbf{b}, Fourier decomposition coefficients directly extracted from tr-SHG data. Temperature is set at 9 K.  \textbf{c}, Fourier coefficients reconstructed from the phenomenological complex fit parameters $A, B, C$ and $D$, using Eqs. (\ref{eq56})–(\ref{eq13}). \textbf{d}, THz is polarized along a node of the SHG-RA pattern (purple arrow). \textbf{e}, Fourier decomposition coefficients directly extracted from tr-SHG data. \textbf{f}, Fourier coefficients reconstructed from the phenomenological complex fit parameters $A, B, C$ and $D$, using Eqs. (\ref{eq56})–(\ref{eq13}).}
    \label{fourier_decomp}
\end{figure*}

\subsection*{SHG intensity derivation: symmetry-allowed tensor components}
Although the atomic positions in MnPS$_3$ are centrosymmetric, the onset of magnetic order breaks inversion symmetry, enabling a $c$-type SHG response~\cite{wang2024,chu2020,ni2021}. Below the N\'eel temperature $T_N$, the magnetic point group reduces to $2'/m$, equivalent to monoclinic $m = C_{1h}$ \cite{shen1984,boyd2008}. Assuming the mirror plane lies in the $ac$ plane, the non-vanishing elements of the second-order susceptibility tensor $\hat\chi$ include:
\[
\chi_{xxx}, \chi_{xyy}, \chi_{xzz}, \chi_{xxz}, \chi_{xzx}, \chi_{yyz}, \chi_{yzy}, \chi_{yxy}, \chi_{yyx}, \chi_{zxx}, \chi_{zyy}, \chi_{zzz}, \chi_{zzx}, \chi_{zxz}.
\]

Symmetry constraints further impose that $\chi_{xxy} = \chi_{xyx} = \chi_{yxx} = \chi_{yyy} = 0$. In the contracted 3×6 matrix notation used for in-plane excitation, the effective susceptibility tensor becomes:
\[
\hat{\chi} = \begin{pmatrix}
\chi_{xxx} & \chi_{xyy} & \chi_{xzz} & \chi_{xyz} & \chi_{xzx} & 0 \\
0 & 0 & \chi_{yzz} & \chi_{yyz} & \chi_{yzx} & \chi_{yxy} \\
\chi_{zxx} & \chi_{zyy} & \chi_{zzz} & \chi_{zyz} & \chi_{zzx} & \chi_{zxy}
\end{pmatrix}.
\]

We consider a normally incident electric field polarized at an angle $\varphi$ with respect to the $y$-axis
\[
\vec{E} = \begin{pmatrix} E_x \\ E_y \\ E_z \end{pmatrix}= E_0\begin{pmatrix} \sin(\varphi) \\ \cos(\varphi) \\ 0 \end{pmatrix}.
\]

For convenience, we use $E_0=1$. The components of second-order polarization $\vec{P}(2\omega)$ can be written as, 
\begin{equation}
    P_i=\sum_{j,k}\chi_{ijk}E_jE_k.
\end{equation}
where $i, j, k \in \{x, y, z\}$. Using the above relation, the SHG polarization is expressed as
\begin{align}
\vec{P}(2\omega) = \begin{pmatrix} P_x \\ P_y \\ P_z \end{pmatrix}=\hat{\chi} \begin{pmatrix} E_x^2 \\ E_y^2 \\ E_z^2 \\ 2E_yE_z \\ 2E_zE_x \\ 2E_xE_y \end{pmatrix}=\begin{pmatrix}
\chi_{xxx} & \chi_{xyy} & \chi_{xzz} & \chi_{xyz} & \chi_{xzx} & 0 \\
0 & 0 & \chi_{yzz} & \chi_{yyz} & \chi_{yzx} & \chi_{yxy} \\
\chi_{zxx} & \chi_{zyy} & \chi_{zzz} & \chi_{zyz} & \chi_{zzx} & \chi_{zxy}
\end{pmatrix} \begin{pmatrix} \sin^2(\varphi) \\ \cos^2(\varphi) \\ 0 \\ 0 \\ 0 \\ 2\sin(\varphi)\cos(\varphi) \end{pmatrix}.
\label{polarization_SHG}
\end{align}

From (\ref{polarization_SHG}), we obtain
\begin{align}
P_x &= \chi_{xxx} \sin^2(\varphi) + \chi_{xyy} \cos^2(\varphi), \\
P_y &= 2\chi_{yxy}\sin(\varphi) \cos(\varphi)=(\chi_{yxy} + \chi_{yyx}) \sin(\varphi) \cos(\varphi), \\
P_z &= \chi_{zxx} \sin^2(\varphi) + \chi_{zyy} \cos^2(\varphi) + (\chi_{zxy}+\chi_{zyx}) \sin(\varphi) \cos(\varphi).
\end{align}

Here, we used the convention $\chi_{ijk}=\chi_{ikj}$. Assuming the detector is aligned with the incident polarization direction
\[
\hat{e}_{\mathrm{det}} = \begin{pmatrix} \sin(\varphi) \\ \cos(\varphi) \\ 0 \end{pmatrix},
\]
the detected SHG signal is thus the projection of the SHG polarization $\vec{P}(2\omega)$ along the detector direction $\hat{e}_{\mathrm{det}}$ which gives
\begin{align}
\vec{P}(2\omega) \cdot \hat{e}_{\mathrm{det}} &= P_x \sin(\varphi) + P_y \cos(\varphi) \\
&= \chi_{xxx} \sin^3(\varphi) + \chi_{xyy} \cos^2(\varphi) \sin(\varphi) + (\chi_{yxy}+\chi_{yyx}) \cos^2(\varphi) \sin(\varphi) \\
&= \chi_{xxx} \sin^3(\varphi)  + (\chi_{xyy}+\chi_{yxy}+\chi_{yyx}) \cos^2(\varphi) \sin(\varphi).
\end{align}

Therefore, the SHG intensity becomes
\begin{align}
I_{SH} = |\vec{P}(2\omega) \cdot \hat{e}_{\mathrm{det}}|^2 = \Big|\; &\chi_{xxx} \sin^3(\varphi)  + (\chi_{xyy}+\chi_{yxy}+\chi_{yyx}) \cos^2(\varphi) \sin(\varphi) \;\Big|^2\\
&= \left| A \cos^2(\varphi) \sin(\varphi) + B \sin^3(\varphi) \right|^2,
\label{SHG_symm}
\end{align}
where $A = \chi_{xyy} + \chi_{yxy} + \chi_{yyx}$ and $B = \chi_{xxx}$. We note that Eq. (\ref{SHG_symm}) resembles Eq. (\ref{fit_1}), providing a full derivation of the phenomenological fitting function for the equilibrium SHG-RA patterns.

\subsection*{SHG intensity derivation: lower symmetry case}
In the photoinduced case, we consider a symmetry-lowered scenario where all $\chi_{ijk}$ elements may be nonzero. The susceptibility tensor is expressed as
\[
\hat{\chi} = \begin{pmatrix}
\chi_{xxx} & \chi_{xyy} & \chi_{xzz} & \chi_{xyz} & \chi_{xzx} & \chi_{xxy} \\
\chi_{yxx} & \chi_{yyy} & \chi_{yzz} & \chi_{yyz} & \chi_{yzx} & \chi_{yxy} \\
\chi_{zxx} & \chi_{zyy} & \chi_{zzz} & \chi_{zyz} & \chi_{zzx} & \chi_{zxy}
\end{pmatrix}.
\]

Following the similar mathematical operations discussed earlier, we obtain the SHG polarization
\begin{align}
P_x &= \chi_{xxx} \sin^2(\varphi) + \chi_{xyy} \cos^2(\varphi) + (\chi_{xxy}+\chi_{xyx}) \sin(\varphi) \cos(\varphi), \\
P_y &= \chi_{yxx} \sin^2(\varphi) + \chi_{yyy} \cos^2(\varphi) + (\chi_{yxy}+\chi_{yyx}) \sin(\varphi) \cos(\varphi).
\end{align}

The detected signal can be written as
\begin{align}
\vec{P}(2\omega) \cdot \hat{e}_{\mathrm{det}} &= P_x \sin(\varphi) + P_y \cos(\varphi) \\
&= \chi_{xxx} \sin^3(\varphi) + \chi_{xyy} \cos^2(\varphi) \sin(\varphi) + (\chi_{xxy}+ \chi_{xyx})\sin^2(\varphi) \cos(\varphi) \\
&\quad + \chi_{yxx} \sin^2(\varphi) \cos(\varphi) + \chi_{yyy} \cos^3(\varphi) + (\chi_{yxy}+\chi_{yyx}) \cos^2(\varphi) \sin(\varphi).
\end{align}

Now, we compute the SHG intensity
\begin{equation}
\begin{aligned}
I_{SH} = |\vec{P}(2\omega) \cdot \hat{e}_{\mathrm{det}}|^2 = \Big|\; &\chi_{xxx} \sin^3(\varphi) + (\chi_{xyy} + \chi_{yxy}+\chi_{yyx}) \cos^2(\varphi) \sin(\varphi) \\
&+ (\chi_{yxx} + \chi_{xxy}+ \chi_{xyx}) \sin^2(\varphi) \cos(\varphi) + \chi_{yyy} \cos^3(\varphi) \;\Big|^2
\end{aligned}
\end{equation}
\begin{equation}
= \left| A \cos^2(\varphi) \sin(\varphi) + B \sin^3(\varphi) + C \sin^2(\varphi) \cos(\varphi) + D \cos^3(\varphi) \right|^2.
\label{SHG_low_symm}
\end{equation}

Here, $C= \chi_{yxx} + \chi_{xxy}+ \chi_{xyx}$ and $D= \chi_{yyy}$. The SHG intensity expression in Eq. (\ref{SHG_low_symm}) mirrors Eq. (\ref{fit_2}). This correspondence confirms that the fitting parameters $C$ and $D$ effectively capture the dynamical symmetry modulations to the SHG-RA patterns as observed experimentally. 

\subsection*{THz pump -- magneto-optic probe measurements}
\begin{figure}
    \centering
    \includegraphics[width=\linewidth]{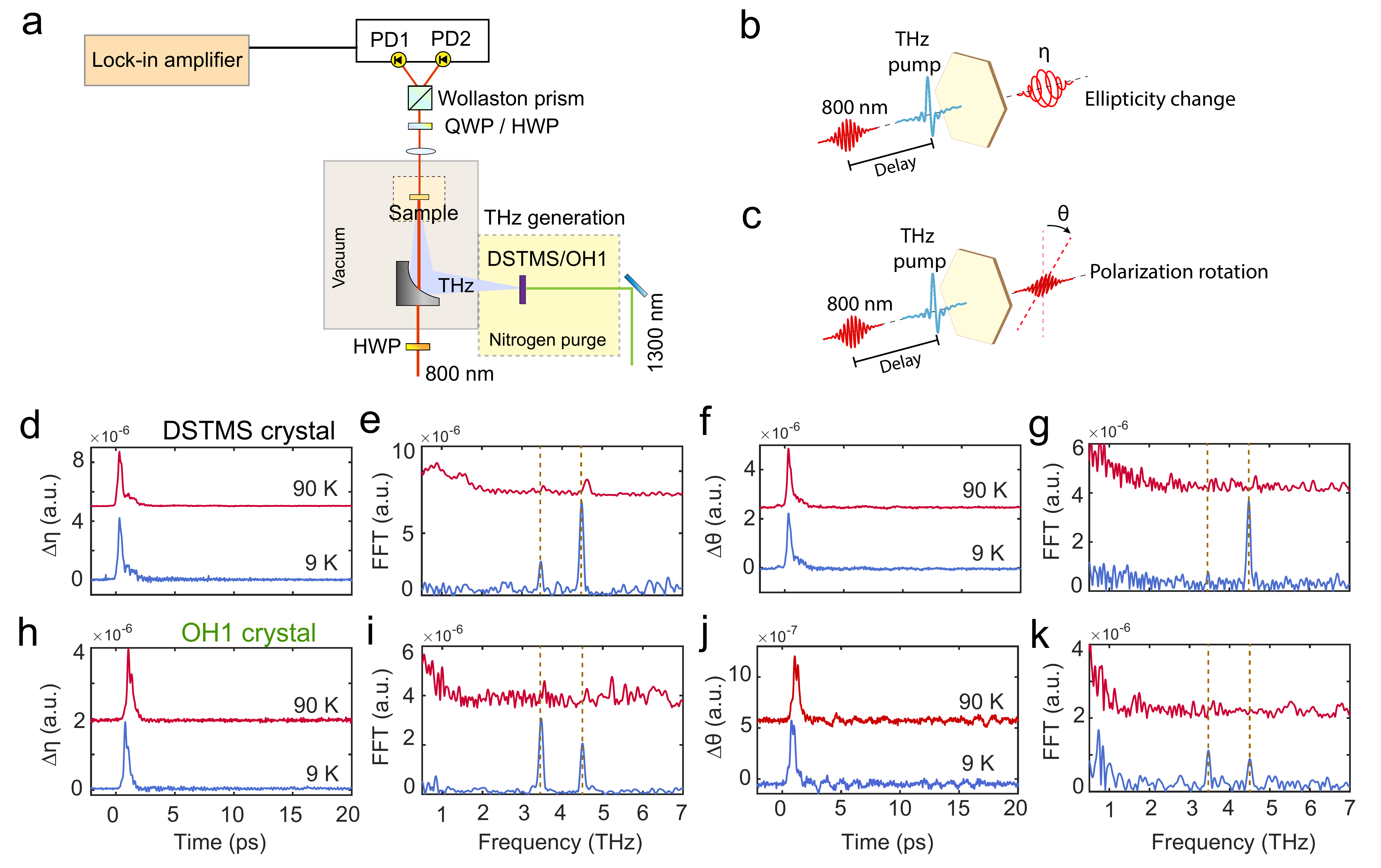}
    \caption{\textbf{THz pump-induced ellipticity change and polarization rotation dynamics.} \textbf{a}, Experimental configuration. HWP: half-wave plate, QWP: quarter-wave plate, PD: photodiode. The QWP is used for measuring ellipticity change, while the HWP is used for polarization rotation measurements. \textbf{b--c}, Schematic illustrations of the measurement geometry. The THz pump is polarized along a lobe of the static SHG-RA pattern. \textbf{d}, Temporal evolution of THz-induced ellipticity change $\Delta \eta$ at temperatures below (9~K, blue) and above (90~K, red) $T_N$. Traces are vertically offset for clarity. DSTMS crystal is used for THz generation. \textbf{e}, Corresponding Fourier transform spectra, revealing coherent Raman-active modes at 3.5 THz and 4.5~THz (brown dashed lines). Although weaker in amplitude, both modes exhibit a blueshift above $T_\mathrm{N}$. \textbf{f--g}, Temperature-dependent dynamics of THz-induced polarization rotation $\Delta \theta$ and its Fourier spectrum. \textbf{h--i}, $\Delta \eta$ dynamics and corresponding FFT spectra at 9~K and 90~K (OH1 as the THz generation crystal), showing the 3.5~THz and 4.5~THz modes. \textbf{j--k}, $\Delta \theta$ dynamics and FFT spectra under the same pumping conditions.}
    \label{ellipticity_polrot}
\end{figure}

Using the same laser source as tr-SHG measurements, we performed dynamic magneto-optic experiments under THz excitation. These measurements, which track changes in the polarization state of a transmitted 800 nm probe, include both THz-induced ellipticity change ($\Delta \eta$) and polarization rotation ($\Delta \theta$). A quarter-wave plate (QWP) and a half-wave plate (HWP) were used to measure $\Delta \eta$ and $\Delta \theta$, respectively.  In both configurations, a Wollaston prism was placed after the waveplates in the beam path, splitting the beam into orthogonally-polarized components to be detected by two Thorlabs DET100A Si-based balanced detectors (Supplementary Fig. \ref{ellipticity_polrot}a-c). We note that for each temperature and experimental configuration, 60 individual measurements were acquired and averaged to improve the signal-to-noise ratio.

The time trace of THz-induced ellipticity change is shown in Supplementary Fig. \ref{ellipticity_polrot}d at below (9K, blue) and above (90 K, red) the $T_N$. A DSTMS crystal was used for THz generation. The signals are vertically offset for clarity. A sharp peak appears near time zero which is followed by fast oscillations. The peak feature is ascribed to THz Kerr effect. Fourier transform of the oscillatory signal at 9 K reveals two Raman-active phonons at 3.5 THz and 4.5 THz (Supplementary Fig. \ref{ellipticity_polrot}e). At 90 K, both modes exhibit a blueshift and a marked reduction in amplitude, consistent with prior Raman measurements \cite{sun2019}, and indicative of their coupling to the AFM order. In addition, temperature-dependent THz-induced polarization rotation signals are shown in Supplementary Fig. \ref{ellipticity_polrot}f, with the associated FFT spectra presented in Supplementary Fig. \ref{ellipticity_polrot}g. While the signal amplitude is notably weaker compared to the ellipticity response, both phonon modes are still resolved at 9 K. The phonon features also diminish in strength and exhibit a blueshift at 90 K, in tandem with the ellipticity change data.

Similar measurements were performed using a THz pump generated from an OH1 crystal (Supplementary Fig.~\ref{ellipticity_polrot}h–k), yielding results consistent with the DSTMS data. The same Raman-active phonons at 3.5 THz and 4.5 THz were observed in both ellipticity change and polarization rotation channels, with comparable temperature-dependent behavior. 

\clearpage
\subsection*{Symmetry modulation analysis when THz is polarized along a node}
\begin{figure*}[hbt!]
    \centering
    \includegraphics[width=\linewidth]{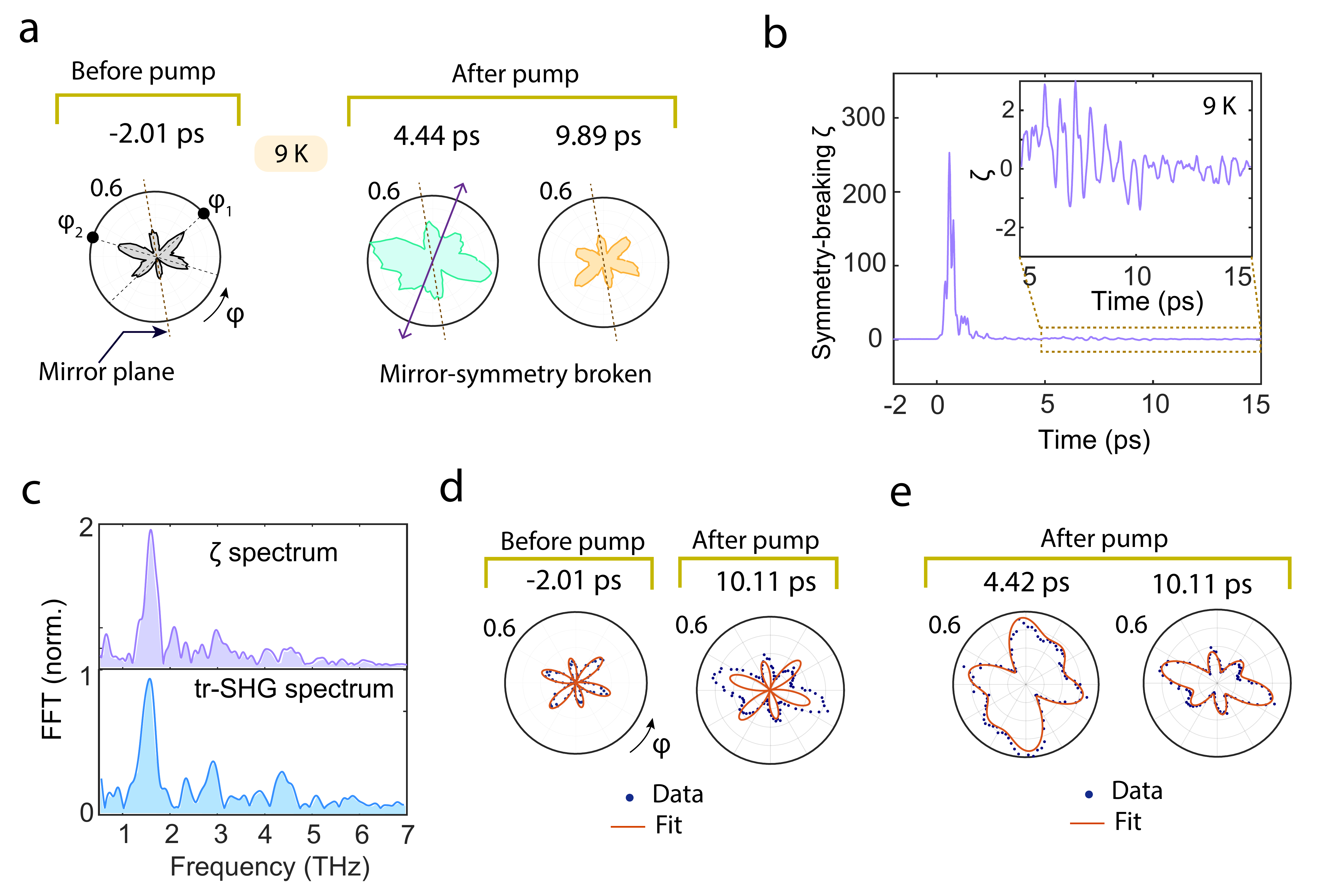}
    \caption{\textbf{THz field-induced long-lived mirror symmetry modulations when THz is along a node (DSTMS crystal). a}, SHG-RA patterns in equilibrium (black), and at different pump-probe time delays upon THz excitation (teal and yellow) at 9 K. Photoexcited SHG-RA patterns exhibit a notable symmetry modulation relative to the equilibrium mirror plane (brown dashed line). The THz pump field is polarized along a node (purple arrow). \textbf{b}, Time-evolution of symmetry-breaking parameter $\zeta$, unmasking coherent oscillations at later times (inset). \textbf{c}, Fourier transform of $\zeta$ (top), compared with tr-SHG spectrum (bottom), demonstrating peaks at 1.7 THz and 4.5 THz. \textbf{d}, Equilibrium SHG-RA pattern at $t = -2.01$ ps fitted with complex parameters $A$ and $B$; this model does not reproduce the photoinduced patterns after THz excitation (i.e., $t = 10.11$ ps). \textbf{e}, Close agreement between nonequilibrium data and phenomenological fitting using complex parameters $A, B, C$ and $D$. The radial scale of each polar plot is normalized, with the outer circle corresponding to an amplitude of 0.6.}
    \label{symm_break_node}
\end{figure*}

\clearpage
\subsection*{Time-resolved SHG dynamics above the AFM transition temperature}
\begin{figure*}[hbt!]
    \centering
    \includegraphics[width=\linewidth]{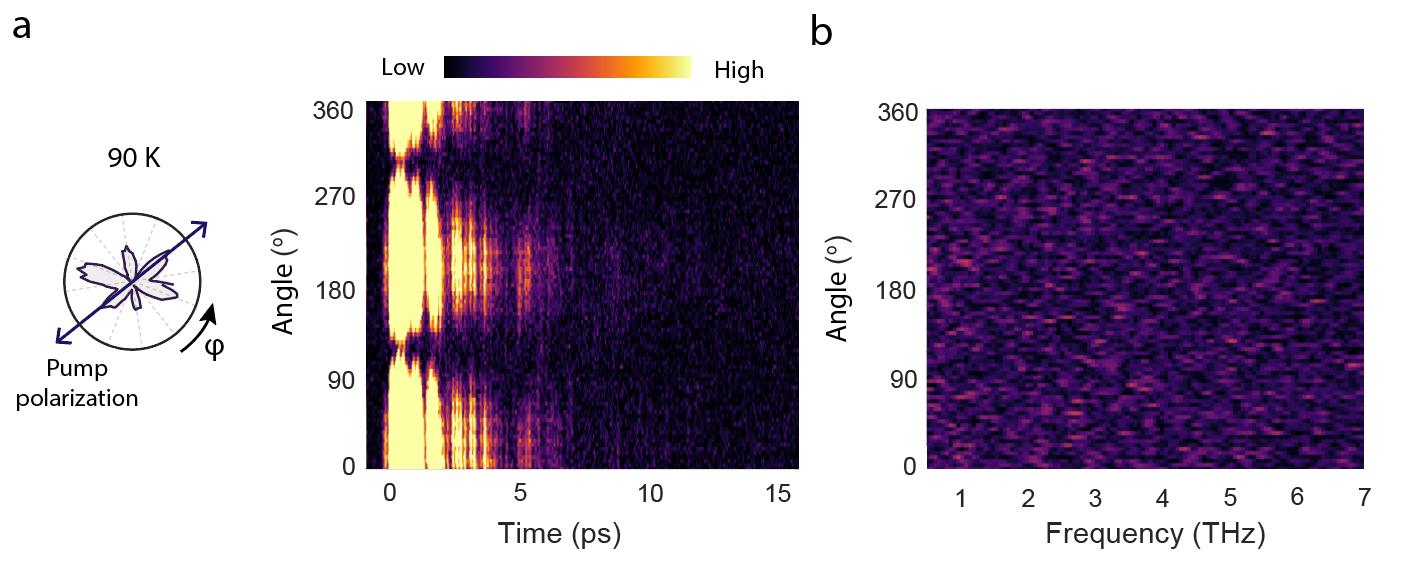}
    \caption{\textbf{Time- and angle-resolved SHG dynamics upon strong-field THz excitation at 90 K. a}, Pump-induced SHG as a function of pump-probe delay time $t$ and polarizer analyzer angle $(\varphi)$ at 90 K. THz pump generated from DSTMS crystal with field strength of $\sim$ 500 kV/cm is polarized along a lobe of the static SHG-RA pattern (dark blue arrow). The THz is generated from a DSTMS crystal. The signal is dominated by TFISH effect with no coherent oscillations. \textbf{c}, Fourier transform spectrum of the pump-probe signal, no coherent modes are detected. }
    \label{fig:90K}
\end{figure*}

\clearpage
\subsection*{Dynamics with THz pump generated from OH1 crystal}
\begin{figure} [hbt!]
    \centering
    \includegraphics[width=\linewidth]{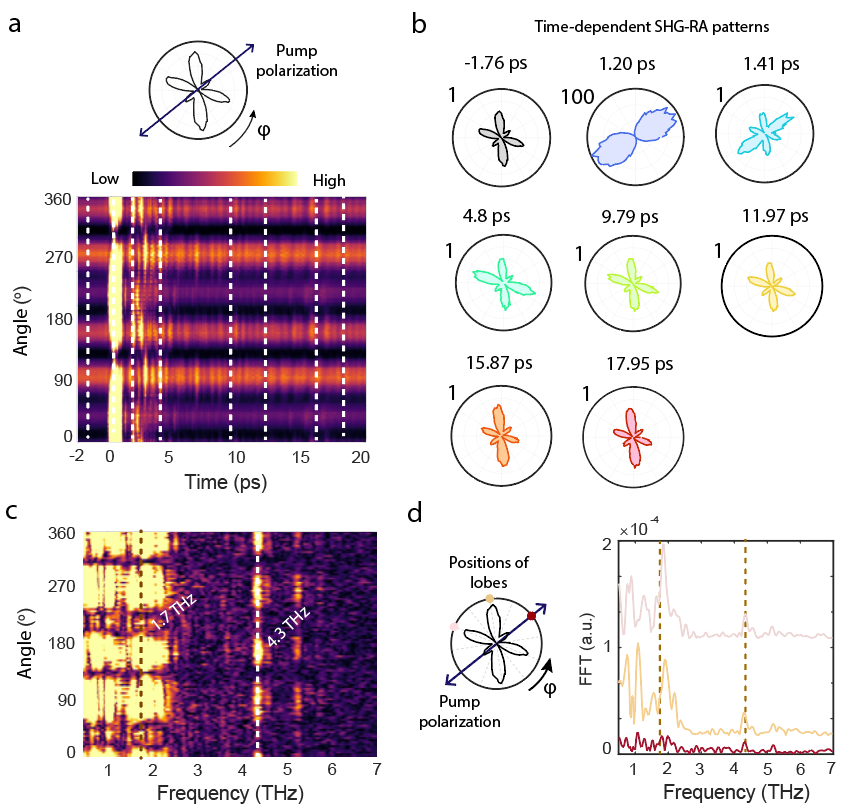}
    \caption{\textbf{Time- and angle-resolved SHG dynamics upon strong-field THz excitation (OH1 crystal). a}, Pump-induced SHG as a function of pump-probe delay time $t$ and polarizer analyzer angle $(\varphi)$ at 9 K. THz pump with field strength of $\sim$ 200 kV/cm is polarized along a lobe of the static SHG-RA pattern (dark blue arrow). The later time dynamics is dominated by a long-lived oscillatory signal, suggesting coherent phonons. \textbf{b}, SHG-RA patterns at different pump-probe delays, obtained from vertical linecuts in \textbf{a} (dashed white lines). The radial scale of each polar plot is normalized, with the outer circle corresponding to an amplitude of 1. \textbf{c}, Fourier transform spectrum of the pump-probe signal, multiple prominent peaks are observed. \textbf{d}, Horizontal linecuts showing FFT spectra at lobe positions (indicated by colored circles in the SHG-RA pattern), showing coherent modes predominantly at 1.7 THz and 4.3 THz. The linecuts are vertically offset for better visibility.}
    \label{OH1_dynamics}
\end{figure}

\clearpage
\subsection*{Field dependence of the coherent modes when THz is generated from OH1 crystal}
\begin{figure} [hbt!]
    \centering
    \includegraphics[width=\linewidth]{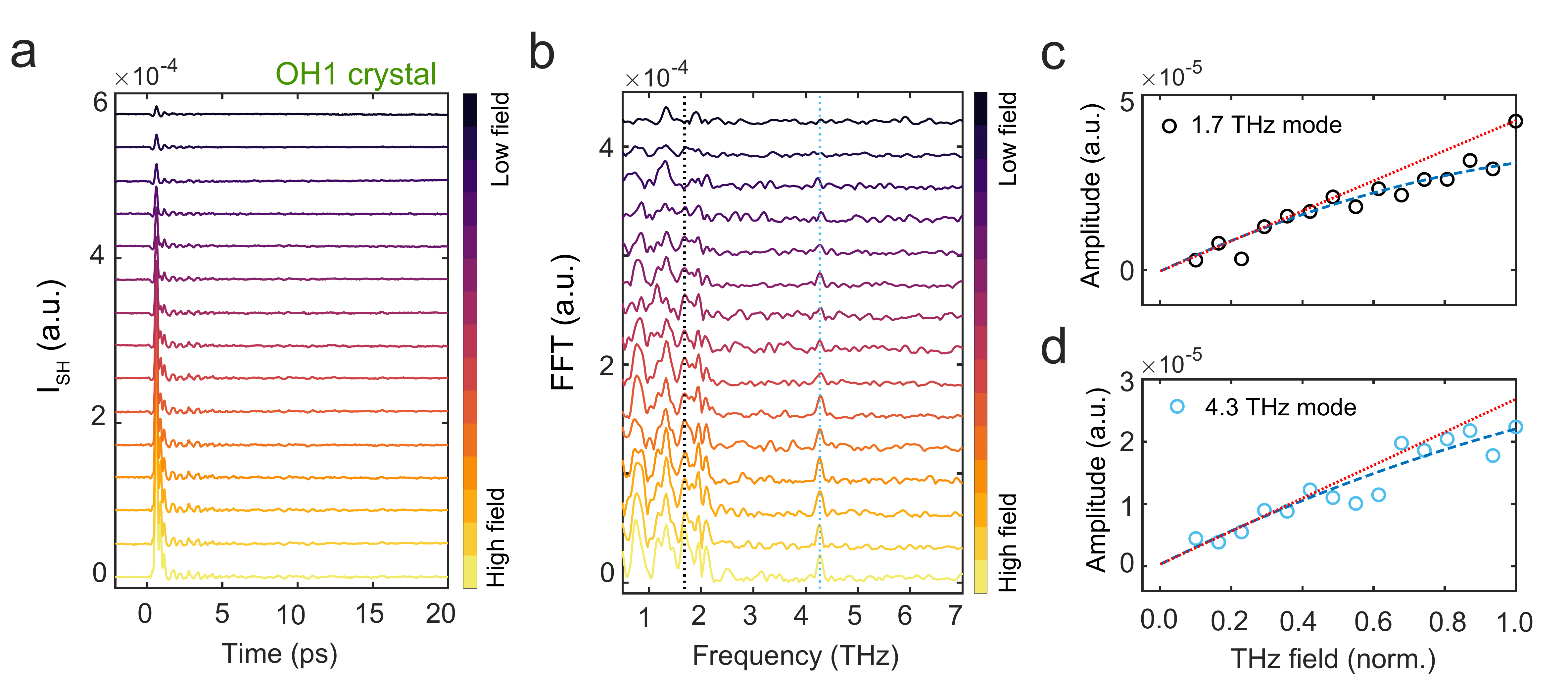}
    \caption{\textbf{tr-SHG dynamics as a function of THz pump field. a}, The tr-SHG dynamics along a lobe as a function of THz field generated from a OH1 crsytal (peak field $\sim 200$ kV/cm) at 9 K. \textbf{b}, Corresponding FFT spectra showing pronounced peak features at 1.7 THz and 4.3 THz (vertical dashed lines). The signals are vertically offset for clarity. \textbf{c--d}, Mode amplitude plotted as a function of normalized pump field for 1.7 THz (\textbf{c}, black open circles) and 4.3 THz (\textbf{d}, blue open circles) modes. Both modes exhibit predominantly linear scaling with field. Red and blue dashed lines indicate linear and sublinear fits, respectively.}
    \label{field_dependence_OH1}
\end{figure}

\clearpage
\subsection*{Symmetry modulation analysis when THz is generated from OH1 crystal}
\begin{figure} [hbt!]
    \centering
    \includegraphics[width=\linewidth]{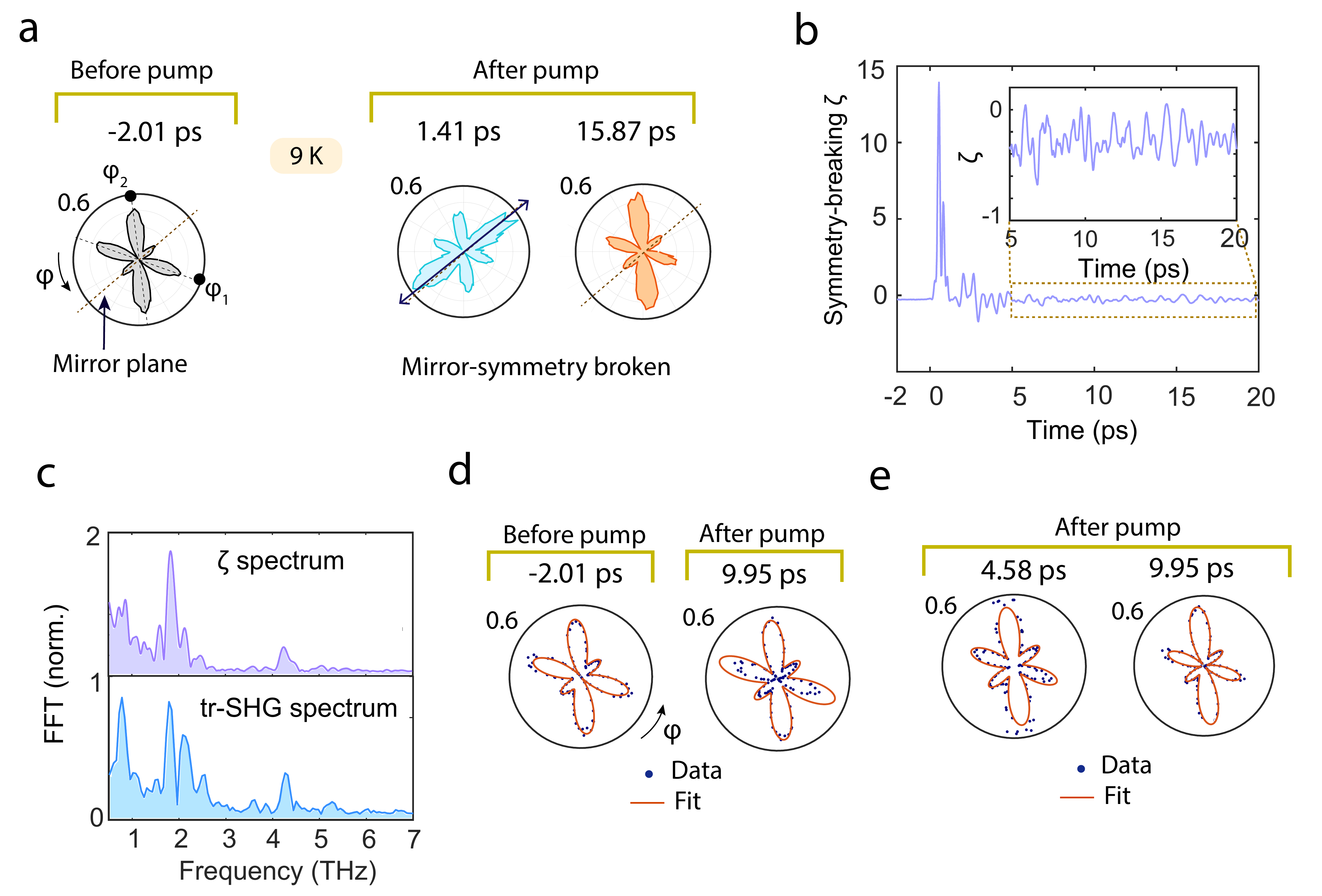}
    \caption{\textbf{Terahertz field-induced long-lived mirror symmetry modulations (OH1 crystal). a}, SHG-RA patterns in equilibrium (black), and at different pump-probe time delays upon THz excitation (light blue and orange) at 9 K. Photoexcited SHG-RA patterns exhibit a notable symmetry modulation relative to the equilibrium mirror plane (brown dashed line). The THz pump field is polarized along a lobe. \textbf{b}, Time-evolution of symmetry-breaking parameter $\zeta$, unmasking coherent oscillations at later times (inset). \textbf{c}, Fourier transform of $\zeta$ (top), compared with tr-SHG spectrum (bottom), demonstrating peaks at 1.7 THz and 4.3 THz. \textbf{d}, Equilibrium SHG-RA pattern at $t = -2.01$ ps fitted with complex parameters $A$ and $B$; this model does not reproduce the photoinduced patterns after THz excitation (i.e., $t = 9.95$ ps). \textbf{e}, Close agreement between nonequilibrium data and phenomenological fitting using complex parameters $A, B, C$ and $D$. The radial scale of each polar plot is normalized, with the outer circle corresponding to an amplitude of 0.6.}
    \label{OH1_symm}
\end{figure}

\clearpage
\subsection*{Dynamics of phenomenological fitting parameters (OH1 generation crystal)}
\begin{figure} [hbt!]
    \centering
    \includegraphics[width=\linewidth]{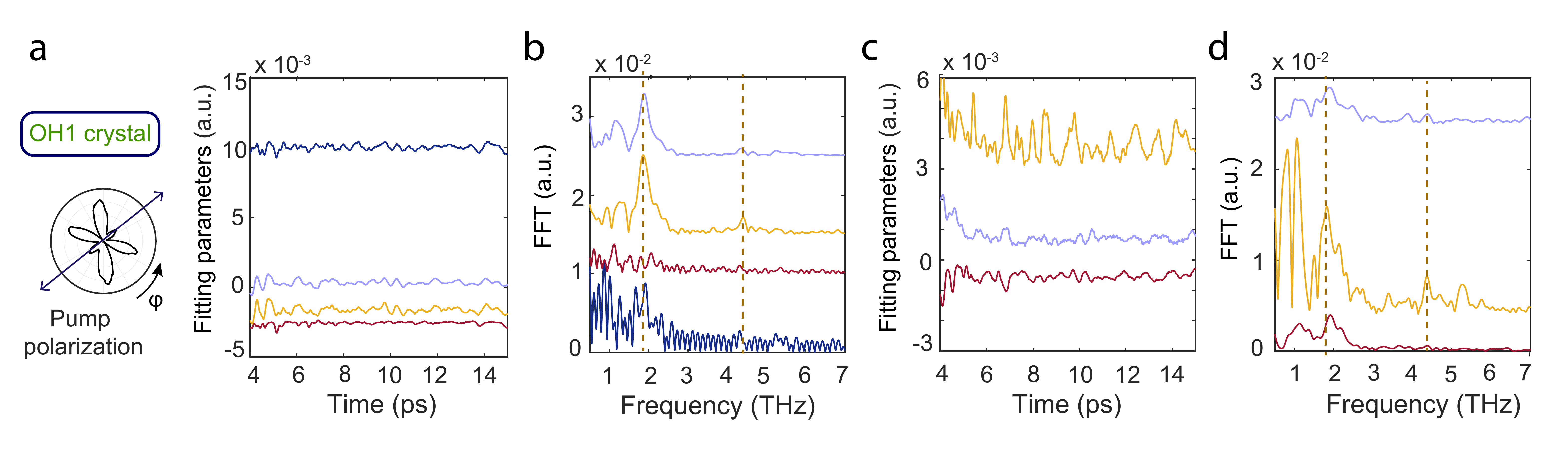}
    \caption{\textbf{Time evolution of complex fitting parameters for the SHG-RA patterns. a--d}, Real (\textbf{a}) and imaginary (\textbf{c}) parts of the SHG fit parameters $A$, $B$, $C$, and $D$, and their corresponding Fourier transforms (\textbf{b}, \textbf{d}) at 9 K. The THz pump is generated from an OH1 crystal, and is polarized along a lobe relative to the SHG-RA pattern (dark blue arrow). Vertical dashed lines at 1.7~THz and 4.3~THz highlight dominant coherent modes. The FFTs are separated by vertical offsets for clarity.}
    \label{fitting_OH1}
\end{figure}

\clearpage
\subsection*{THz-induced modification of bond strength derived from tight-binding model calculation}
\begin{figure} [hbt!]
    \centering
    \includegraphics[width=\linewidth]{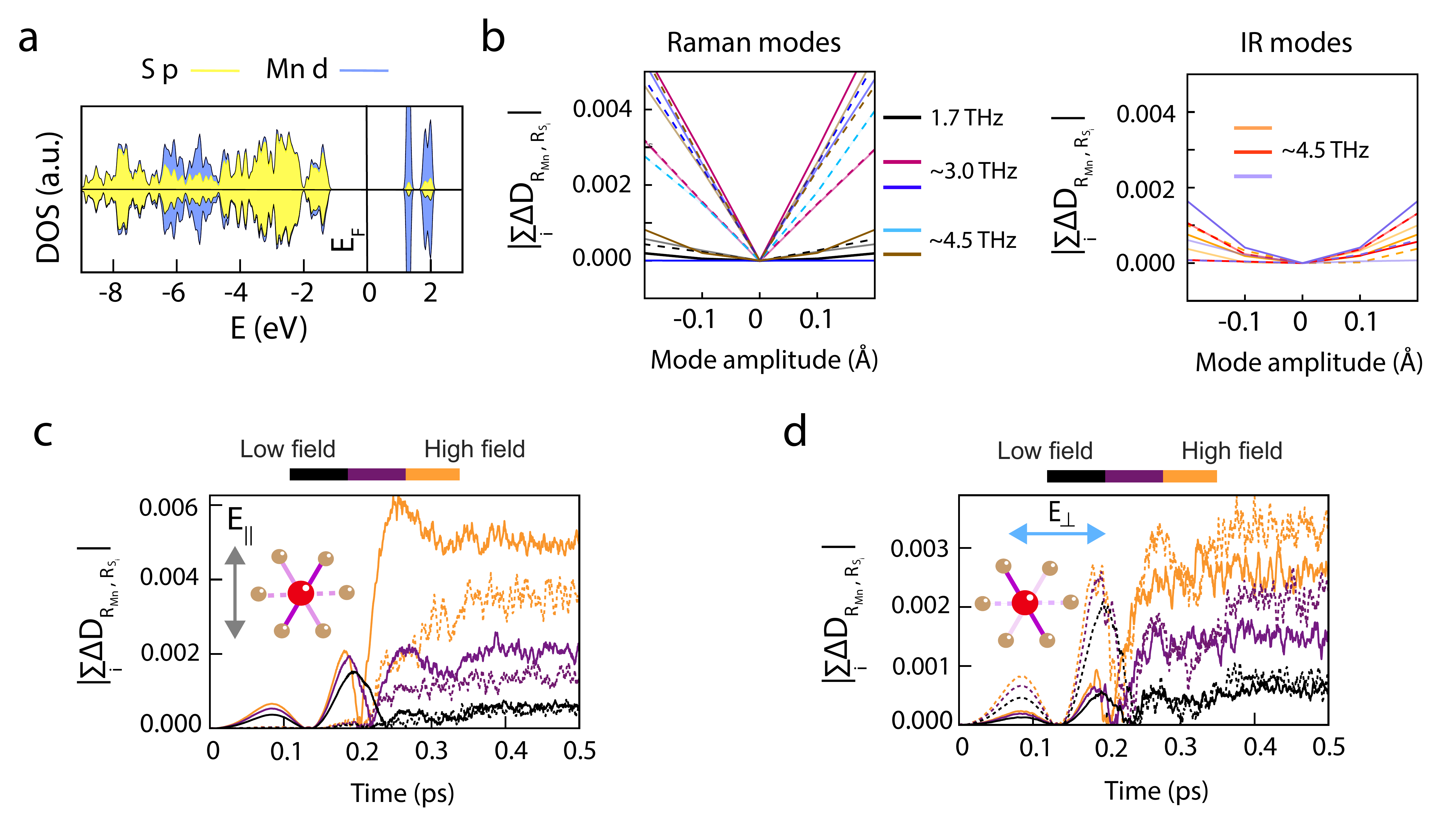}
    \caption{\textbf{Tight-binding model-based calculations.} \textbf{a}, Density of states of the ground-state of MnPS$_3$ calculated from the tight-binding model projected on S $p $ (yellow) and Mn $d $ orbitals. Projected DOS are stacked on top of each other. The two spin channels are shown with opposite signs. The THz pump field is polarized along a lobe. \textbf{b}, Phonon induced modifications in Mn--S bond strength around Mn atoms. Solid, dashed, and transparent lines represent Mn–S bonds oriented along three different directions. \textbf{c--d}, Evolution of Mn--S bonding strength $D_{R_{\mathrm{Mn}},R_{S_i}}$ under the THz field from the model-based rt-TDDFT simulations for $E_\parallel$ (\textbf{c}) and $E_\perp$ (\textbf{d}) polarizations. Solid, dashed, and transparent lines denote Mn–S bonds in three different directions around Mn atoms, as shown in the inset.}
    \label{tb_model}
\end{figure}

\clearpage
\subsection*{Constrained DFT simulation of 4.5 THz Raman mode driving under charge transfer condition}
\begin{figure} [hbt!]
    \centering
    \includegraphics[width=80 mm]{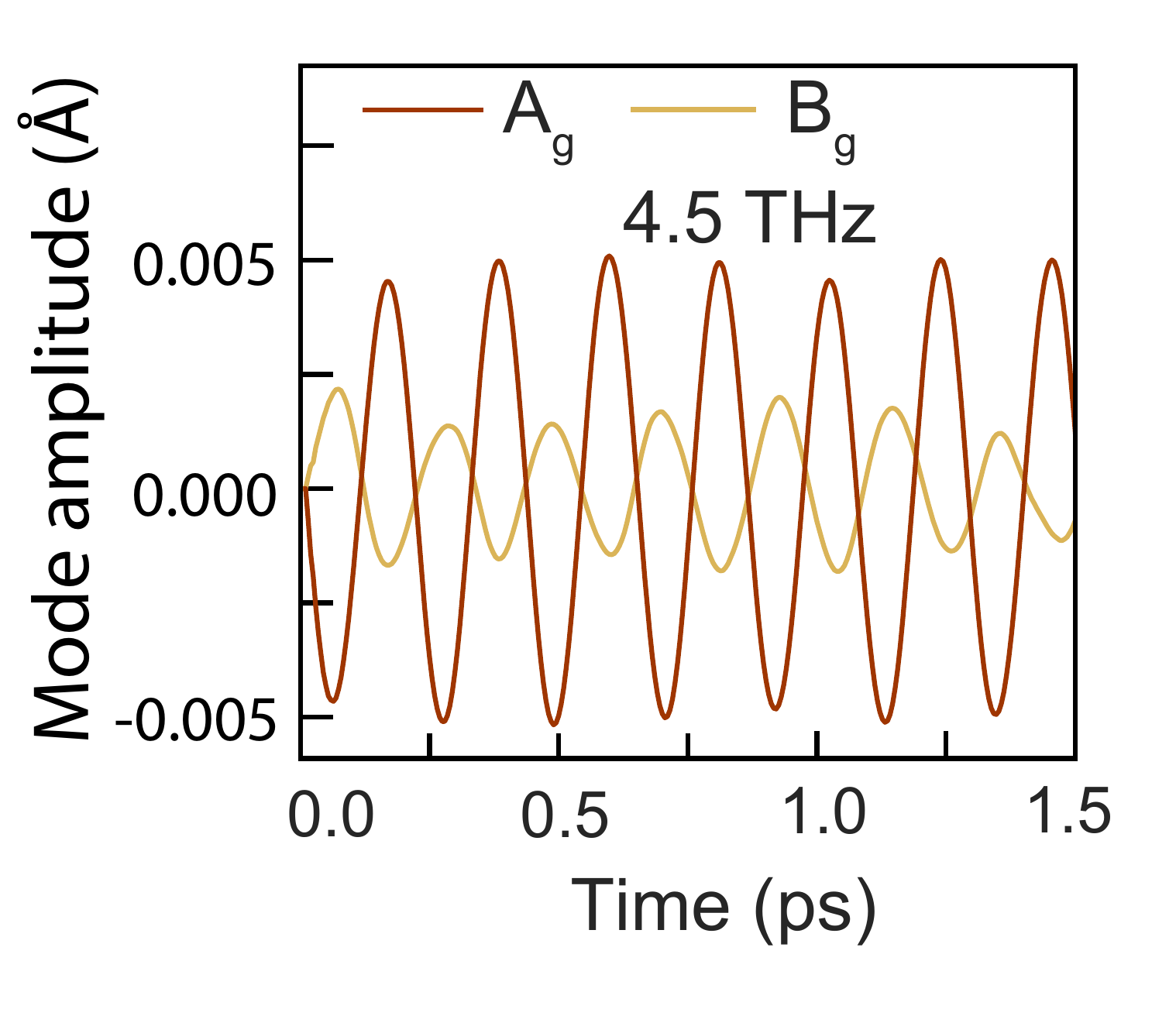}
    \caption{\textbf{Simulation of 4.5 THz Raman mode under THz-induced charge transfer.} Molecular dynamics simulations with a fixed electronic structure and a 0.2\% excited state population, initialized with atomic velocity distribution corresponding to $T=9$ K, revealing the dynamics of the 4.5 THz Raman modes with $A_g$ (maroon) and $B_g$ (beige) symmetries.}
    \label{4d5_constrained}
\end{figure}

\clearpage
\subsection*{Movies showing temporal evolution of SHG-RA patterns}

We provide four movies visualizing the time evolution of SHG-RA patterns under strong-field THz excitation (peak field strength $\sim500$ kV/cm) at 9 K, using a DSTMS crystal for THz generation:

\begin{itemize}
    \item \textbf{Movie S1:} THz field is polarized along a lobe (equivalent to $E_\perp$) of the static SHG-RA pattern, displayed with a fixed radial scale (capped at $1 \times 10^{-5}$) to enable direct temporal comparison.
    
    \item \textbf{Movie S2:} THz field is polarized along a node (equivalent to $E_\parallel$), also shown with a fixed radial scale for consistent comparison across delay times.
    
    \item \textbf{Movie S3:} Same configuration as Movie S1 (THz field polarized along a lobe), but using a dynamic radial scale that adjusts to the signal magnitude in each frame.
    
    \item \textbf{Movie S4:} Same configuration as Movie S2 (THz field polarized along a node), also displayed with a dynamic radial scale.
\end{itemize}

All four movies are provided as individual \texttt{.mp4} files.

\subsection*{Movies showing dynamical phenomenological fitting}

Two additional movies present time-resolved phenomenological fitting of SHG-RA patterns at 9 K under THz excitation:

\begin{itemize}
    \item \textbf{Movie S5:} THz field is polarized along a lobe. Experimental SHG-RA patterns (solid blue circles) are fitted at each time step using the expression in Eq.~(\ref{SHG_low_symm}) in Supplementary Materials, with red curves representing the phenomenological fits using complex fitting parameters $A$, $B$, $C$, and $D$. For better visibility, the fixed radial scale is capped at $8 \times 10^{-6}$.
    
    \item \textbf{Movie S6:} THz field is polarized along a node, with experimental data and fitting displayed in the same format as in Movie S5.
\end{itemize}

These fitting dynamics are also provided as separate \texttt{.mp4} files.
\end{document}